\documentclass[natbib]{svjour3}                     
\smartqed  
\usepackage{graphicx}
\def\simgreat{\lower2pt\hbox{$\buildrel {\scriptstyle >}
   \over {\scriptstyle\sim}$}}
\def\simless{\lower2pt\hbox{$\buildrel {\scriptstyle <}
   \over {\scriptstyle\sim}$}}

\def\msun{$M_{\odot}$}

\begin{document}

\title{Black Hole Spin via Continuum Fitting and the Role of Spin in
  Powering Transient Jets}


\author{Jeffrey E. McClintock  \and Ramesh Narayan \and James F. Steiner
}


\institute{J. E. McClintock, R. Narayan, and J. F. Steiner \at
              Harvard-Smithsonian Center for Astrophysics, 60 Garden Street, Cambridge, MA 02138 USA\\
              email:~{jem@cfa.harvard.edu; narayan@cfa.harvard.edu; jsteiner@cfa.harvard.edu}           
}

\date{Received: 2013 March 3 / Accepted: 2013 June 24}

\maketitle

\begin{abstract}
The spins of ten stellar black holes have been measured using the
continuum-fitting method.  These black holes are located in two
distinct classes of X-ray binary systems, one that is persistently
X-ray bright and another that is transient.  Both the persistent and
transient black holes remain for long periods in a state where their
spectra are dominated by a thermal accretion disk component.  The spin
of a black hole of known mass and distance can be measured by fitting
this thermal continuum spectrum to the thin-disk model of Novikov and
Thorne; the key fit parameter is the radius of the inner edge of the
black hole's accretion disk.  Strong observational and theoretical
evidence links the inner-disk radius to the radius of the innermost
stable circular orbit, which is trivially related to the dimensionless
spin parameter $a_*$ of the black hole ($|a_*| < 1$).  The ten spins
that have so far been measured by this continuum-fitting method range
widely from $a_*\approx0$ to $a_*>0.95$.  The robustness of the method
is demonstrated by the dozens or hundreds of independent and
consistent measurements of spin that have been obtained for several
black holes, and through careful consideration of many sources of
systematic error.  Among the results discussed is a dichotomy between
the transient and persistent black holes; the latter have higher spins
and larger masses.  Also discussed is recently discovered evidence in
the transient sources for a correlation between the power of ballistic
jets and black hole spin.

\end{abstract}


\section{Introduction}\label{sec:intro}

In his Ryerson Lecture, \citet{cha+1975} described the Kerr solution
as the ``most shattering experience'' of his entire scientific life.
He found himself ``shuddering before the beautiful, the incredible
fact'' that each of the many trillions of black holes in the universe
is completely described by a single pair of numbers that specify the
black hole's mass and its spin \footnote{Spin is commonly expressed in
terms of the dimensionless parameter $a_* \equiv cJ/GM^2$, where
$J$ and $M$ are respectively the angular momentum and mass of the
black hole.}.  In Chandrasekhar's time, 1910--1995, the masses of Cyg
X-1 and three other stellar black holes had been estimated
\citep{web+1972,bol+1972,cow+1983,mcc+1986,rem+1992,sha+1994}.  Today,
accurate dynamical mass measurements have been achieved for more than
a dozen stellar black holes
\citep{mcc_rem+2006,oze+2010,oro+2011a,stg+2013}, as well as for
several supermassive black holes, e.g., Sgr A*
\citep{ghe+2008,gil+2009}, NGC 4258 \citep{her+2005}, and others
\citep[][and references therein]{gul+2009}.

In 1989, the first practical approach to measuring black hole spin was
suggested by \citet{fab+1989}, namely, modeling the
relativistically-broadened Fe K emission line emitted from the inner
accretion disk.  The first compelling observation of such a line was
reported just two months before Chandrasekhar died \citep{tan+1995}.
Presently, the spins of more than a dozen black holes have been
estimated by modeling the ``reflected'' spectrum of an accretion disk,
which includes as its most prominent feature the Fe K line.  For a
review of this method of measuring black hole spin, we refer the
reader to \citet{rey+2013}.

It was not until 1997 that a new approach to measuring black hole spin
-- the continuum-fitting method -- which is the subject of this
chapter, was pioneered by \citet{zha+1997}.  In brief, in applying
this method one fits the thermal continuum spectrum of a black hole's
accretion disk to the relativistic thin-disk model of \citet{nov+1973}
and thereby determines the radius of the inner edge of the disk.  One
then identifies this radius with the radius of the innermost stable
circular orbit ($R_{\rm ISCO}$), which is simply related to the spin
parameter $a_*$ \citep{bar+1972}.  The method is simple: It is
strictly analogous to measuring the radius of a star whose flux,
temperature and distance are known.  By this analogy, it is clear that
it is essential to know the luminosity of the accretion disk; hence,
one must have estimates of the source distance $D$ as well as the disk
inclination $i$.  Additionally, one must know $M$ in order to scale
$R_{\rm ISCO}$ and thereby determine $a_*$.

In 2006, the continuum-fitting method was employed to estimate the
spins of three stellar black holes \citep{sha+2006,mcc+2006}.
Presently, ten spins have been measured using this method
(Section~\ref{sec:results}).  Not only is the continuum-fitting method
simple, it is also demonstrably robust.  For example, there is strong
observational and theoretical evidence (discussed in
Section~\ref{sec:isco}) that the disk is truncated quite sharply at
$R_{\rm ISCO}$.  Furthermore, there is an abundance of suitable X-ray
spectral data for many black holes; consequently, for a given black
hole one can typically obtain tens or even hundreds of independent
measurements of spin that agree to within a few percent
(Section~\ref{sec:isco_obs}).  The one open question for this method
is whether the black hole's spin is aligned with the orbital angular
momentum vector of the inner disk (Section~\ref{sec:error_align}).
Meanwhile, a limitation of the continuum-fitting method is that it is
only readily applicable to stellar black holes \citep[but
see][]{jol+2009,cze+2011}, while the Fe K method is applicable to both
stellar and supermassive black holes \citep{rey+2013}.

In order to obtain secure measurements of spin using the
continuum-fitting method, and to establish the reliability of this
method, substantial and comparable effort is required on three fronts:
(1) The selection and fitting of X-ray spectral data to the
Novikov-Thorne model (in conjunction with ancillary models); (2)
testing and exploring extensions of the Novikov-Thorne model via
general relativistic magnetohydrodynamic (GRMHD) simulations; and (3)
obtaining accurate estimates of $D$, $i$ and $M$.  The first two
topics are discussed in Sections~\ref{sec:cf}--\ref{sec:error}.
Concerning the third topic, we refer the reader to recent papers on
the measurements of these crucial parameters for M33 X-7
\citep{oro+2007}; LMC X-1 \citep{oro+2009}; A0620--00
\citep{can+2010}; XTE J1550--564 \citep{oro+2011b}; Cyg X-1
\citep{rei+2011,oro+2011a}; H1743--322 \citep{ste+2012a}; and
GRS~1915+105 \citep{stg+2013}.  The uncertainties in $D$, $i$ and $M$
are critically important because they dominate the error budget in the
final determination of $a_*$, including the error incurred by reliance
on the Novikov-Thorne disk model (Section~\ref{sec:error}).

Initial efforts are under way to use the available continuum-fitting
spin data to investigate the formation and evolution of black holes,
as well as their host systems \citep[e.g.,][]{lee+2002,won+2012}, and
to understand how a black hole interacts with its environment
\citep[e.g.,][]{wan+2003,coo+2008}.  The most important application to
date of spin data is the discovery of a long-predicted correlation
between jet power and black hole spin, which is the subject of
Section~\ref{sec:jets}.  Very recently, \citet{rus+2013} challenged
the validity of this correlation; Section~\ref{sec:jets_challenge}
answers this challenge.


\section{Stellar Black Holes in X-ray Binaries}\label{sec:bhbs}

\begin{figure}\sidecaption
  \includegraphics[width=0.65\textwidth]{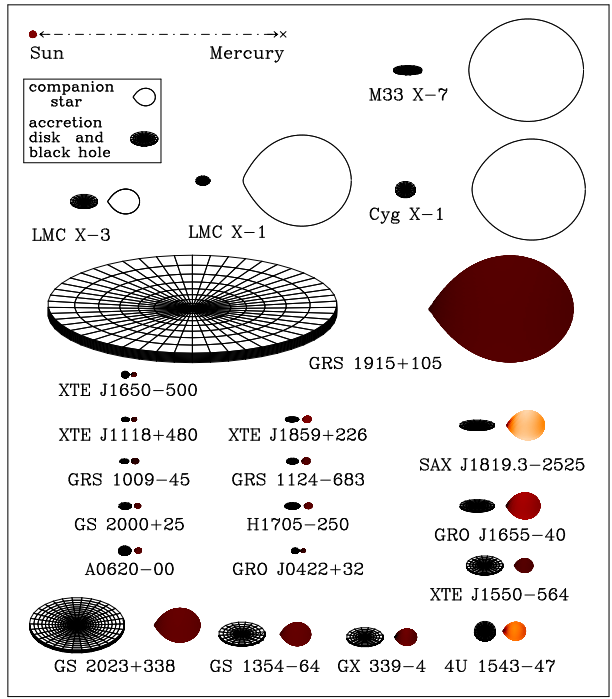}
\caption{Schematic sketch to scale of 21 black hole binaries (see
scale and legend in the upper-left corner).  The tidally-distorted
shapes of the companion stars are accurately rendered in Roche
geometry.  The black holes are located in the center of the disks.  A
disk's tilt indicates the inclination angle $i$ of the binary, where
$i=0$ corresponds to a system that is viewed face-on; e.g.,
$i=21^{\circ}$ for 4U 1543--47 (bottom right) and $i = 75^{\circ}$ for
M33 X-7 (top right).  The size of a system is largely set by the
orbital period, which ranges from 33.9 days for the giant system
GRS~1915+105 to 0.2 days for tiny XTE J1118+480. Three well-studied
persistent systems (M33 X-7, LMC X-1 and Cyg X-1) are located in the
upper-right corner.  The other 18 systems are transients.  (Figure
courtesy of J.\ Orosz.)}
\label{fig:oroszplot}  
\end{figure}

There are 24 confirmed black hole binaries: the 23 listed in Table 1
in \citet{oze+2010} plus H1743--322 \citep{ste+2012a}\footnote{Apart
from H1743--322, our selection is based on firm dynamical evidence,
and we therefore exclude some important systems for which there is
significant evidence that the primary is a black hole, e.g., Cyg~X-3
\citep{zdz+2013}, or a strong presumption that it is, e.g., SS433
\citep{beg+2006} and 4U~1957+11 \citep{now+2012}}.  A schematic sketch
to scale of 21 of these confirmed black-hole systems is shown in
Figure~\ref{fig:oroszplot}.

For decades, it has been customary to define two classes of X-ray
binaries, commonly referred to as LMXBs (low-mass X-ray binaries) and
HMXBs (high-mass X-ray binaries), based on whether the mass of the
secondary star is relatively low or relatively high
\citep[e.g.,][]{bra+1983}.  Here, we use a different classification
scheme that differentiates two distinct classes of black hole binaries
by the primary mode of mass transfer to the black hole and the effect
that this has on the stability of the X-ray source \citep{whi+1995}.

The black holes in five of the 24 systems are steadily fed by the
winds of massive O-supergiant or Wolf-Rayet companions, and
consequently their bolometric X-ray luminosities are relatively
stable.  Sketches of three of these systems (M33 X-7, LMC X-1 and Cyg
X-1) appear in the top-right corner of Figure~\ref{fig:oroszplot}. We
refer to these systems and their black holes as ``persistent.''

The black holes in the remaining systems are fed by Roche lobe
overflow through the L1 point, and all of them have been observed to
vary in luminosity by factors of $>100$ \citep[$\sim10^8$ in several
extreme cases; e.g., see][]{nar+2008}.  We refer to these systems and
their black holes as ``transient.''

\subsection{Persistent Black Hole Binaries}\label{sec:bhbs_persistent}

These systems are distinguished by the large masses of their secondary
stars ($20~M_{\odot}-70~M_{\odot}$) and by the extreme optical/UV
luminosities of these stars, which exceed the X-ray luminosities of
their black hole companions.  Consequently, the effects of X-ray
heating are minimal and the optical star dominates the optical
properties of the system.  The key distinguishing feature of these
systems is their X-ray persistence, which is a consequence of the
star's massive stellar wind ($\sim10^{-5}-10^{-8}
M_{\odot}~$yr$^{-1}$), a significant fraction of which is captured by
the black hole.

Because the secondaries are massive these systems are obviously young
($\simless~10^{7}$yr).  They are also very rare: There is only one
confirmed system in the Galaxy, Cyg X-1, and, despite many deep {\it
Chandra} and {\it XMM-Newton} X-ray observations of Local Group
galaxies, only four other such systems have been discovered, one each
in the LMC, M33, IC~10 and NGC~300.

In this review, we do not consider further the two persistent systems
that contain Wolf-Rayet secondaries, namely IC~10~X-1 and NGC~300~X-1,
because the masses of their black holes depend strongly on the very
uncertain masses of their secondaries, and also because no attempt has
so far been made to estimate their spins.  By contrast, the three
remaining persistent systems -- M33 X-7, LMC X-1 and Cyg X-1 --
have well-determined values of both mass and spin (see
Section~\ref{sec:results}).  Relative to the black hole primaries in
the transient systems (apart from GRS~1915+105), the black holes in
these three persistent systems have large masses, $M =
11-16~M_{\odot}$, and high spins that range from $a_* = 0.84$ to $a_*
> 0.95$, a point that we return to in
Section~\ref{sec:results_compare}.

\subsection{Transient Black Hole Binaries}\label{sec:bhbs_transient}

\begin{center}
\begin{figure}\sidecaption
  \includegraphics[width=0.68\textwidth]{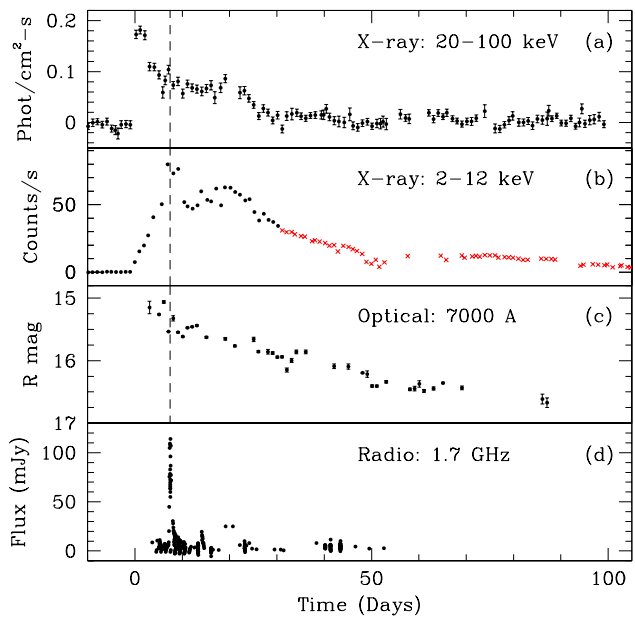}
\caption{Outburst cycle of XTE~J1859+226 in 1999.  The dashed line
(top three panels) marks the time of peak radio flux (panel d).  The
$\approx1$-day radio spike (panel d) is shown fully resolved in
Figure~2 in \citet{bro+2002}.  The red crosses (panel b) indicate
times when the X-ray spectrum is dominated by the thermal component.
These {\it BATSE} and {\it RXTE}/ASM X-ray, and Merlin (and other)
radio data (panels a, b and d, respectively) appear in Figure~1 in
\citet{bro+2002} and the optical data (panel c) appear in Figure~2 in
\citet{san+2001}.  For further details, consult the references.}
\label{fig:j1859}       
\end{figure}
\end{center}

With few exceptions, the 18 transient black hole binaries (hereafter
simply referred to as transients) manifest and then rise to maximum
X-ray luminosity on a timescale of several days, thereafter returning
to a quiescent state over a period of many tens or hundreds of days,
as illustrated in Figure~\ref{fig:j1859}.  The masses of the black
holes in these systems are relatively low, as are their spins (with
the exception of GRS~1915+105), and their orbital periods range widely
from 0.2--33.9~days.  By comparison, the orbital periods of the
persistent systems span a relatively narrow range.  The transients
are, on average, likely Gyrs old \citep{whi+1998,fra+2013}.

During a major outburst, the peak luminosities of transient sources
approach the Eddington limit \citep{ste+2013}, while in quiescence
their luminosities are typically in the range $10^{-8.5}$ to $10^{-6}$
of Eddington \citep{nar+2008}.  Figure~\ref{fig:j1859} shows X-ray,
optical and radio light curves of a typical short-period transient.
The optical emission is generated largely by reprocessing of X-rays in
the accretion disk, and the radio outburst is primarily the result of
synchrotron emission produced in a jet.  The ballistic jets, which are
the subject of Section~\ref{sec:jets}, are launched very near the time
of peak radio emission (panel d), which in the case of XTE J1859+226
occurred just 0.5~days after the X-ray luminosity (panel b) peaked.
Spin can be reliably measured when the thermal component dominates the
spectrum during the latter part of the outburst cycle (panel b).  For
a complete and state-coded version of the X-ray light curve of
XTE~J1859+226, see Figure~8b in \citet{rem+2006}.

There are several oddballs among the transient systems: Four have
relatively massive secondaries, $\sim2-6~M_{\odot}$, compared to the
typical value of $\simless~1~M_{\odot}$ \citep{cha+2006}.
GRS~1915+105 has remained very luminous continuously since its
appearance in 1992, and GX 339-4 never reaches a deep quiescent state
\citep{mcc_rem+2006}.  LMC X-3 is almost always active and highly
variable (Section~\ref{sec:isco_obs}), although it does have extended
low states \citep{sma+2012}.


\section{The Continuum-Fitting Method}\label{sec:cf}

The two foundations of the continuum-fitting method are (1) the
existence of an ISCO for a test particle orbiting a black hole and (2)
the strong observational and theoretical evidence that -- for a wide
range of conditions -- accretion disks in black hole binaries are
truncated quite sharply at the ISCO radius.  In this section, we first
discuss the physics of these disks, and we close by describing the
mechanics of continuum fitting.

\subsection{Accretion Disk Theory}\label{sec:cf_theory}
\label{accdisktheory}

The basic physics of black hole accretion is straightforward
\citep{fra+2002,kat+2008,abr+2013}. Gas with angular momentum flows in
from the outside and settles into a circular orbit stabilized by
centrifugal force. The gas steadily loses angular momentum as a result
of magnetic stresses from the magnetorotational instability
\citep{bal+1998}, whose effect is often approximated via the
$\alpha$-viscosity prescription of \citet{sha+1973}.  As the gas loses
angular momentum, it moves inward, occupying at each instant a
circular orbit appropriate to its instantaneous angular momentum. The
inward drift continues until the gas reaches the radius of the ISCO,
$R_{\rm ISCO}$. Inside $R_{\rm ISCO}$, no stable circular orbits are
available and the gas falls dynamically into the black hole.

\begin{figure}
  \includegraphics[width=0.49\textwidth]{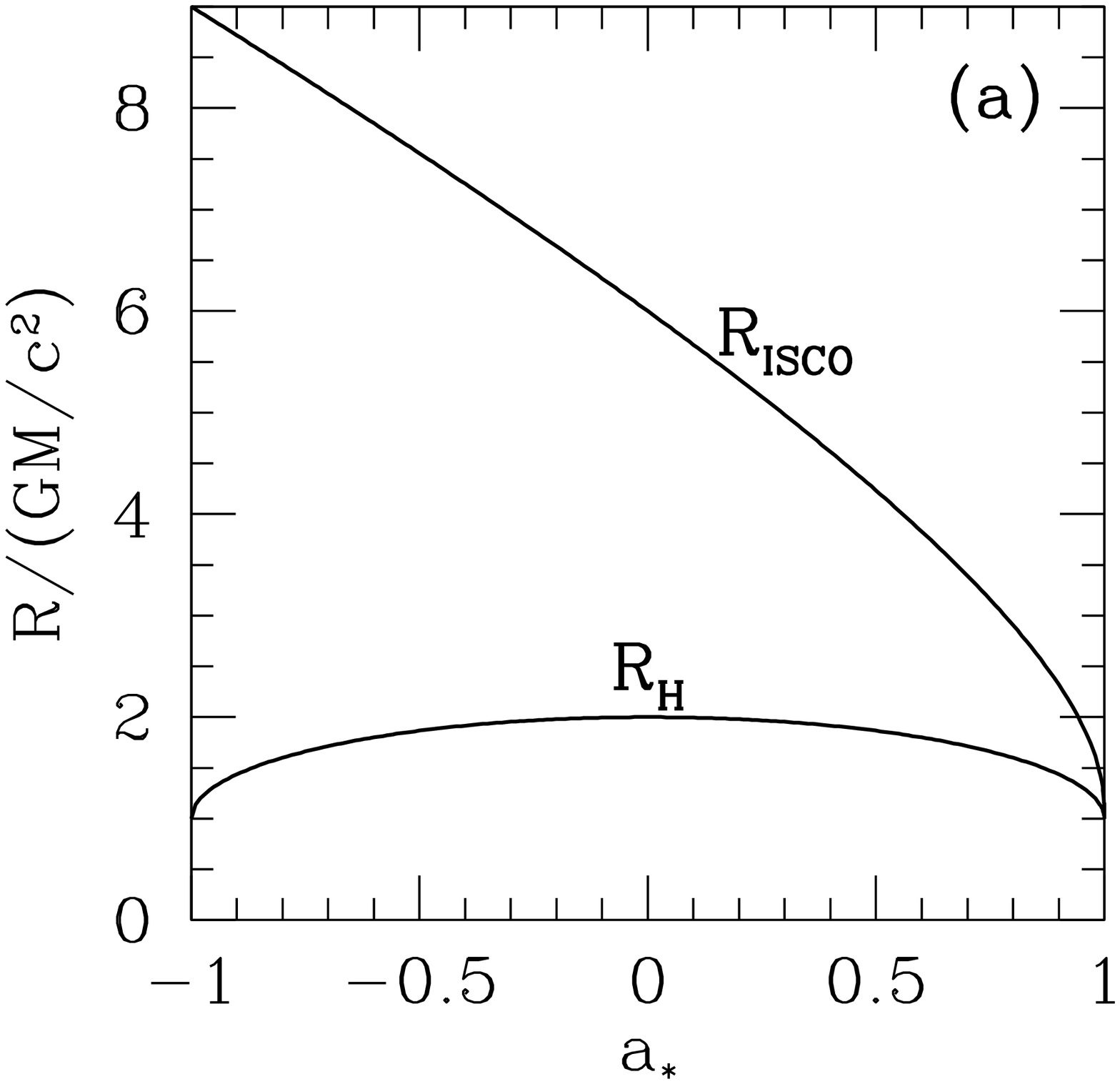}
  \includegraphics[width=0.55\textwidth]{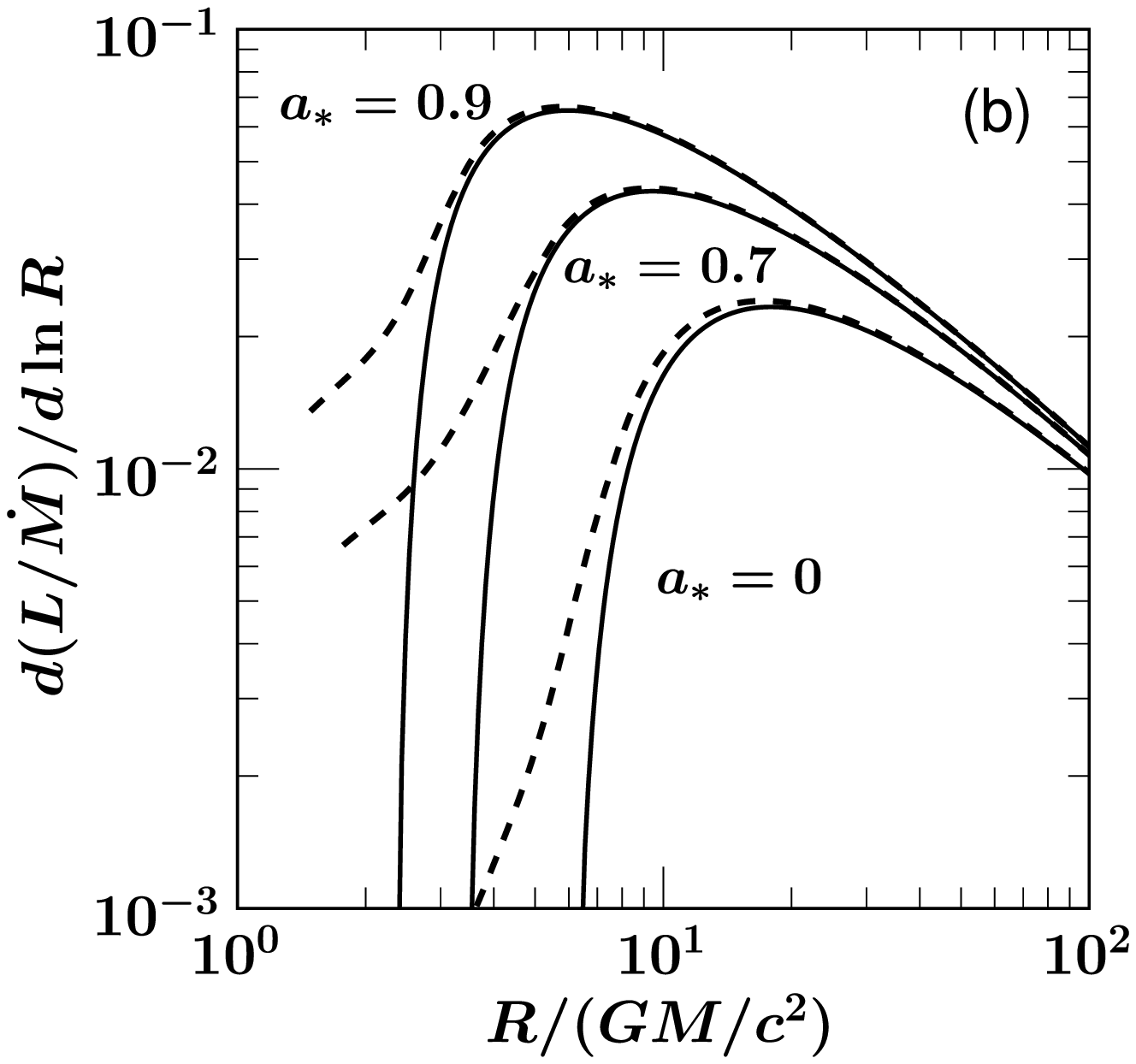}
\caption{(a) Radius of the ISCO $R_{\rm ISCO}$ and of the horizon
  $R_{\rm H}$ in units of $GM/c^2$ plotted as a function of the black
  hole spin parameter $a_*$. Negative values of $a_*$ correspond to
  retrograde orbits. Note that $R_{\rm ISCO}$ decreases monotonically
  from $9GM/c^2$ for a retrograde orbit around a maximally spinning
  black hole, to $6GM/c^2$ for a non-spinning black hole, to $GM/c^2$
  for a prograde orbit around a maximally spinning black hole. (b)
  Profiles of $d(L/\dot{M})/d\ln R$, the differential disk luminosity
  per logarithmic radius interval normalized by the mass accretion
  rate, versus radius $R/(GM/c^2)$ for three values of $a_*$.  Solid
  lines are the predictions of the NT model.  The dashed curves from
  \citet{zhu+2012}, which show minor departures from the NT model, are
  discussed in  Section~\ref{sec:error_NT}.}
\label{fig:RISCO}
\end{figure}

As described above, the ISCO represents a major transition point in
disk physics, where gas switches from slow viscous accretion on the
outside to inviscid free-fall on the inside.  The ISCO is thus
effectively the inner edge of the disk. Correspondingly, information
on the linear dimensions of the radius $R_{\rm ISCO}$ is imprinted on
the emitted radiation.  Since $R_{\rm ISCO}$ varies monotonically with
the black hole spin parameter $a_*$ \citep{bar+1972}, as illustrated
in Figure~\ref{fig:RISCO}a, it is thus possible to measure $a_*$ by
modeling the disk emission.

The model of choice for this purpose is that described by
\citet{nov+1973}, hereafter referred to as the NT model, which is the
relativistic generalization of the thin accretion disk model of
\citet{sha+1973}. Using nothing more than the Kerr metric, basic
conservation laws of mass, momentum, angular momentum and energy, and
assumptions of axisymmetry and steady state, the NT model \citep[see
also][]{pag+1974,rif+1995} derives an analytical formula for the
differential luminosity $dL(R)/dR$ emitted by the disk as a function
of radius $R$.

The solid lines in Figure~\ref{fig:RISCO}b show for three values of
$a_*$ the differential disk luminosity predicted by the NT model. The
disk flux vanishes at $R_{\rm ISCO}$ because the model has, by
assumption, no viscous stress inside this radius (see
Section~\ref{sec:isco_theory} for further discussion).  More
importantly, the peak emission occurs at a radius that tracks the ISCO
(it is a factor of $\sim2-3$ larger than $R_{\rm ISCO}$).  This means
that the radiation is emitted from a progressively smaller effective
area, roughly $\propto R_{\rm ISCO}^2$, as the black hole spin
increases. Therefore, for a given total disk luminosity, the
temperature of the emitted radiation increases with increasing
$a_*$. This is the key physical effect that underlies the
continuum-fitting method.  By measuring the characteristic temperature
and luminosity of the disk emission, and applying the NT model, one is
able to estimate both $a_*$ and the mass accretion rate $\dot{M}$.

As should be clear from the above, the accuracy of the
continuum-fitting method ultimately depends on the reliability of the
NT model; this issue is discussed further in
Sections~\ref{sec:isco_theory} and \ref{sec:error_NT}. It also depends
on our ability to calculate the spectrum of the radiation, which would
be trivial if the disk radiated as a perfect blackbody.
Unfortunately, because electron scattering plays a prominent role at
the X-ray temperatures found in black hole binaries, the emitted
spectrum is substantially harder than a blackbody spectrum of the same
flux. Hence it is necessary to employ detailed disk atmosphere
models. Most of the work to date is based on the atmosphere model {\sc
bhspec} developed by \citet{dav_hub+2006}, which is discussed in
Sections~3.2~and~5.3.

\subsection{Continuum Fitting in Practice}\label{sec:cf_practice}

In broad outline, one fits the X-ray continuum spectrum to the
Novikov-Thorne model of a thin accretion disk with other spectral
components as needed, principally a Compton component.  As stressed in
Sections~1~and~5, in order to obtain useful constraints on $a_*$,
one must inform the fitting process by inputting accurate values of the
external parameters $D$, $i$ and $M$.  The spectral fit returns two
output parameters: the spin $a_*$ and the mass accretion rate $\dot
M$. An important derived quantity is the Eddington-scaled luminosity
of the disk component $L(a_*,\dot M)/L_{\rm Edd}$.

In practice, one usually fits the thermal component using {\sc
kerrbb2} \citep{mcc+2006}\footnote{For alternatives, see
\citet{gie+2001}, \citet{kol+2010}, and \citet{str+2011}.}, which is a
hybrid code implemented in XSPEC \citep{arn+1996} that combines the
capabilities of two relativistic disk models, {\sc bhspec}
{\citep{dav+2005} and {\sc kerrbb} \citep{lil+2005}.  This latter
model, {\sc kerrbb}, which is a straightforward implementation of the
analytic Novikov-Thorne model, has three principal fit parameters:
$a_*$, $\dot M$, and the spectral hardening factor $f$, which relates
the observed color temperature to the effective temperature,
$f=T/T_{\rm eff}$.

In fitting the disk component with {\sc kerrbb}, it is quite generally
the case that one can only determine two parameters, a shape parameter
(e.g., $a_*$ or $T$) and a normalization constant (e.g., $\dot M$).
That is, in practice one cannot additionally obtain a useful
constraint on $f$.  However, this limitation of {\sc kerrbb} is
handily overcome by pairing it with {\sc bhspec}, which is based on
non-LTE disk atmosphere models within an $\alpha$-viscosity
prescription.  {\sc bhspec} has just two principal fit parameters
(spin and mass accretion rate), and it can be used to fit directly for
$a_*$.  However, it does not include the effects of self-illumination
of the disk (``returning radiation''), which is a feature that is
included in {\sc kerrbb}.

The pairing of {\sc kerrbb} and {\sc bhspec} is achieved using {\sc
kerrbb2}, which is a modified version of {\sc kerrbb} that contains a
pair of look-up tables for $f$ corresponding to two values of the
viscosity parameter: $\alpha=0.01$ and 0.1.  The entries in the tables
are computed using {\sc bhspec}.  The two tables give $f$ versus
$L/L_{\rm Edd}$ for a wide range of the spin parameter
($|a_*|\le0.9999$).  The computations of $f$ versus $L/L_{\rm Edd}$
are done using the appropriate, corresponding response matrices and
energy ranges used in fitting the spectra with {\sc kerrbb}.  Thus,
{\sc kerrbb} and the subroutine/table computed using {\sc bhspec}
(which together constitute {\sc kerrbb2}) allow one to fit directly
for $a_*$ and $L/L_{\rm Edd}$, while retaining the returning-radiation
feature of {\sc kerrbb}.

Depending on the quality of a particular spectrum, it may be necessary
to include minor spectral components (e.g., line or edge features),
but these cosmetic features do not significantly affect the spin
results.  Typically, three model components are fitted in conjunction
with the thermal component: a low-energy cutoff, a ``reflected''
component \citep[e.g.,][]{ros+2007}, and a Compton component.  The
cutoff is straightforward to model \citep[e.g.,][]{wil+2000}, and the
reflected component is relatively weak in disk-dominated spectra, even
in the most extreme circumstances \citep{gou+2011}.  It is the
modeling of the Compton component that has been of central concern in
applying the continuum-fitting method, and we discuss this issue now.

All spectra of black hole binaries, even the most disk-dominated, show
a high-energy tail component of emission, which is widely attributed
to Compton upscattering of soft photons by coronal electrons
\citep{rem+2006}.  In early continuum-fitting work
\citep{sha+2006,mcc+2006}, this component was modeled unsatisfactorily
by adding a power-law component to the spectrum.  All subsequent work
has used a much-improved empirical model of Comptonization called {\sc
simpl} \citep{ste+2009b}.  This model self-consistently generates the
Compton component from the thermal seed spectrum of photons.  It
allows reliable measurements of spin to be obtained even as the
fraction of seed photons $f_{\rm SC}$ that are scattered into the
power-law component approaches 25\% \citep{ste+2009a,ste+2009b}.  The
use of {\sc simpl} in place of the standard power law has doubled the
body of useful data for several sources \citep[e.g.,
see][]{ste+2011,ste+2012a}, and it has enabled the measurement of the
spins of black holes whose spectra are persistently quite strongly
Comptonized such as LMC X-1 \citep{gou+2009} and Cyg X-1
\citep{gou+2011}.

Successful application of the continuum-fitting method requires the
selection of spectra that are disk-dominated.  For transient sources,
such spectra are typically observed during the latter part of an
outburst cycle (Figure~\ref{fig:j1859}b).
Figure~\ref{fig:lmcx3_shane} shows a spectrum with a peak flux in the
Compton component that is only 1\% of the peak thermal flux.  For
spectra that are this disk-dominated, how one chooses to model the
adulterating Compton component is obviously quite unimportant.
Meanwhile, there is an abundance of spectra of comparable quality
available for several sources, i.e., sources with $f_{\rm
SC}~\sim~1$\% \citep[e.g., see Figure~1 in][and Table~1 in Steiner et
al.\ 2011]{ste+2009a}.

While it is essential to select spectra that have a substantial
thermal component \citep[i.e., $f_{\rm
SC}~\simless~25$\%;][]{ste+2009a}, it is equally important to select
data of moderate luminosity, specifically spectra with
Eddington-scaled disk luminosities $L/L_{\rm Edd} < 0.3$.  Otherwise,
the disk scale-height grows and the thin disk model is invalidated
\citep[Sections~\ref{sec:isco_theory} and
\ref{sec:error_NT};][]{mcc+2006}.  Fortunately, there is usually an
abundance of such data because a typical transient source remains for
months in a suitable disk-dominated state of moderate luminosity (see
Figure~\ref{fig:j1859}b).  A very wide range of detectors are capable
of providing suitable data (see example in
Section~\ref{sec:isco_obs}).  The principal requirements are that the
data can be corrected for dead time, and that the detector have a
dozen or more energy channels, an appropriate bandwidth, and be well
calibrated (Section~\ref{sec:error_obs}).

\begin{figure}\sidecaption
  \includegraphics[width=0.65\textwidth]{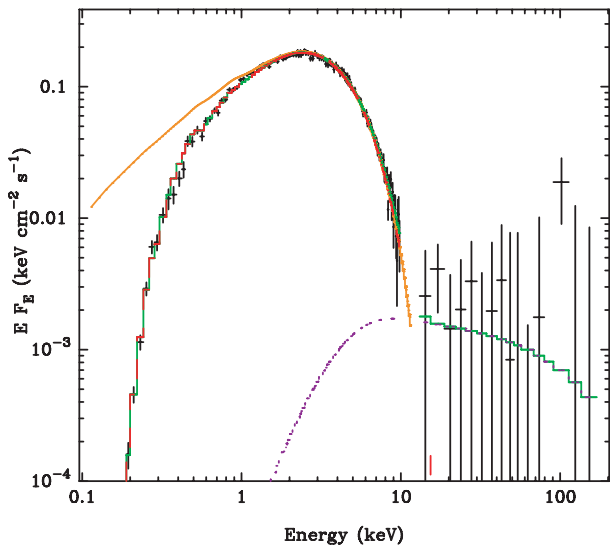}
\caption{Model fit to a disk-dominated spectrum of LMC X-3 obtained
    using detectors aboard the {\it BeppoSAX} satellite for
    $D=52$~kpc, $i=67^{\circ}$ and $M = 10~M_{\odot}$
    \citep{dav+2006}.  A green solid curve, which is difficult to
    discern because it hugs the data, is the total model.  Also shown
    is the thermal component (red long-dashed curve) and the Compton
    component (violet short-dashed curve).  The reflected component is
    negligible and was not included.  The orange solid curve shows the
    total model with the effects of interstellar absorption removed.
    Note that the peak Compton flux is only 1\% of the peak thermal
    flux.}
\label{fig:lmcx3_shane}       
\end{figure}


\section{Truncation of the Disk at the ISCO}\label{sec:isco}

We review the large body of observational evidence that there exists a
constant inner-disk radius in disk-dominated states of black hole
binaries.  We follow with theoretical evidence, based on GRMHD
simulations, that this fixed radius can be identified with the radius
of the ISCO.

\subsection{Observational Evidence}\label{sec:isco_obs}

It has been clear for decades that fitting the X-ray continuum might
prove to be a promising approach to measuring black hole spin.  The
earliest indications came with the advent in the mid-1980s of a
nonrelativistic disk model \citep{mit+1984,mak+1986}, now referred to
as {\sc diskbb}, which returns the color temperature $T_{\rm in}$ at
the inner-disk radius $R_{\rm in}$.  In an important review paper,
\citet{tan_lew+1995} show the remarkable stability of $R_{\rm in}$ for
three transients as the thermal flux of these sources steadily decays
on a timescale of months by factors of 10--100 (see their
Figure~3.14).  Tanaka \& Lewin remark that the constancy of $R_{\rm
in}$ suggests that this fit parameter is related to the radius of the
ISCO.  Subsequently, similar evidence for a constant inner-disk radius
in disk-dominated states of black hole binaries has been demonstrated
for many sources by showing that the bolometric luminosity of the
thermal component is approximately proportional to $T_{\rm in}^4$
\citep{kub+2001,kub+2004,gie+2004,abe+2005,mcc+2009}.

A recent study of the persistent source LMC X-3 presents the most
compelling evidence to date for a constant inner-disk radius
\citep{ste+2010}.  This result is based on an analysis of a large
sample of X-ray spectra collected during eight X-ray missions that
span 26 years.  As illustrated in Figure~\ref{fig:lmcx3_jack} for a
selected sample of 391 {\it RXTE} spectra, the radius of the accretion
disk was found to be constant over time and unaffected by the gross
variability of the source to within $\approx2$~percent.  Even
considering an ensemble of eight X-ray missions, the radius was
observed to be stable to within $\approx5$~percent.  These results
provide compelling evidence for the existence of a fixed inner-disk
radius and establish a firm empirical foundation for the measurement
of black hole spin.  The only reasonable inference is that this radius
is closely associated with the radius of the ISCO, as we show to be
the case in the following section.

\begin{figure}
  \includegraphics[width=1.00\textwidth]{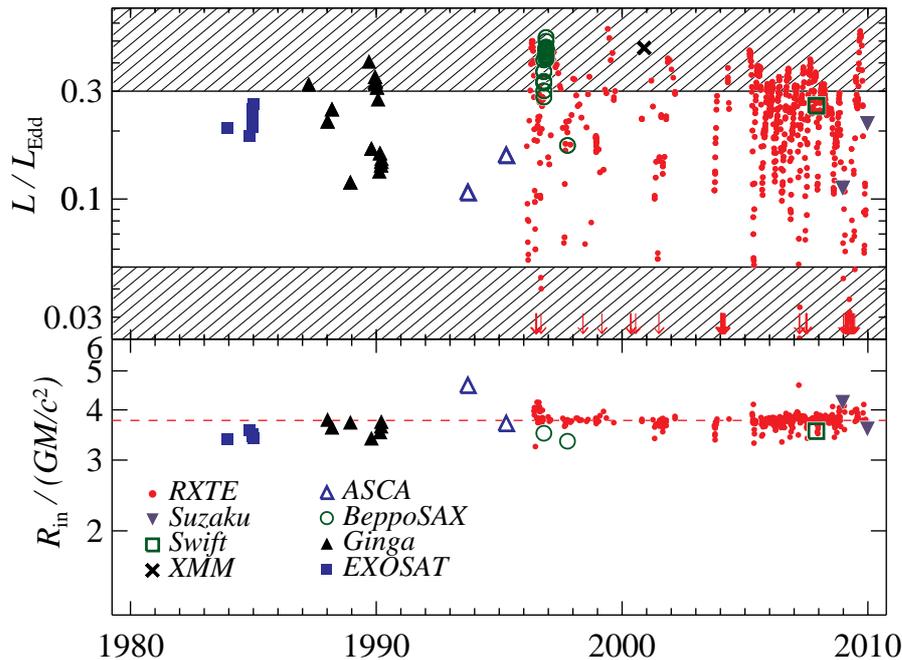}
\caption{$(top)$ Accretion disk luminosity in Eddington-scaled units
(for $M=10$~\msun) versus time for all the 766 spectra considered in a
study of LMC X-3 by \citet{ste+2010}.  (Downward arrows indicate data
that are off scale.)  Selected data in the unshaded region satisfy the
thin-disk selection criterion $L/L_{\rm Edd} < 0.3$ and avoid
confusion with strongly-Comptonized hard-state data with $f_{\rm
SC}~\simgreat~25$\% \citep[Section~3.2;][]{rem+2006}.  $(bottom)$
Fitted values of the inner-disk radius are shown for thin-disk data in
the top panel that meet the selection criteria of the study (a total
of 411 spectra).  Despite large variations in luminosity, $r_{\rm in}$
remains constant to within a few percent over time.  The median value
for just the 391 selected {\it RXTE} spectra is shown as a red dashed
line.}
\label{fig:lmcx3_jack}       
\end{figure}

\subsection{Theoretical Evidence}\label{sec:isco_theory}
\label{RISCOtheory}

The NT model makes one key assumption: It assumes that the viscous
torque vanishes inside the ISCO. While eminently reasonable
\citep[e.g.,][]{pac+2000,afs+2003,sha+2008b}, this ``zero-torque''
assumption does not follow directly from basic conservation laws but
is applied as an extra ad hoc boundary condition. Furthermore, the
luminosity profiles shown for the NT model in
Figure~\ref{fig:RISCO}b depend critically on this boundary condition
because this condition causes the luminosity profiles to vanish at the
ISCO, which in turn fixes the radius of peak disk emission.
\citet{kro+1999} argued that magnetic stresses can operate freely
across the ISCO and will cause strong torques at the ISCO, as well as
in the inner plunging region.  Also, \citet{gam+1999} came up with a
simple analytical MHD model of the plunging region with demonstrably
non-zero torques. What do real disks do?

To answer this question, geometrically thin accretion disks around
black holes have been simulated by a number of authors
\citep{sha+2008a,nob+2009,pen+2010} using state-of-the-art GRMHD
codes. The main advantage of simulations is that they do not require
ad hoc assumptions. One simply introduces magnetized gas in a Kerr
space-time and lets the system evolve to a quasi-steady state.  Since
the ISCO lies inside the simulation box, well away from computational
boundaries, it is not treated differently from other regions of the
system. In other words, no boundary condition is applied by hand at
the ISCO, a great improvement over analytical models. On the other
hand, for technical reasons, simulations to date have not treated
radiation transfer self-consistently but instead have assumed local
cooling. This is not considered serious for the purposes of testing
the zero-torque condition.

The dashed lines in Figure~\ref{fig:RISCO}b show results from
simulations of very thin disks ($H/R\sim0.05$;
\citealt{pen+2010,kul+2011,zhu+2012}).  The simulation-derived disk
luminosity profiles show modest deviations from the NT model
predictions; in particular, the disk flux does not vanish inside the
ISCO.  On the other hand, the deviations are minor, and we discuss and
quantify these effects in Section~\ref{sec:error_NT}.  Importantly,
the radius corresponding to the luminosity peak, which is the most
relevant quantity for the continuum-fitting method, agrees
quite well.  As discussed in Section~\ref{sec:error_NT}, the
good agreement between model predictions and simulation results
translates into modest uncertainties in spin estimates.  GRMHD
simulation results such as these shown in Figure~\ref{fig:RISCO}b are
viewed as a strong validation of the NT model.

Interestingly, the magnetic stress in the simulations does not vanish
in the plunging region. Indeed, \citet{pen+2010} found that the stress
there agrees remarkably well with Gammie's (1999) model. However,
there is little energy dissipation associated with this stress
(Gammie's analytical model has zero dissipation), so it has little
bearing on the continuum-fitting method. Another interesting result is
that deviations from the NT model seem to increase as the luminosity
and, concomitantly, the disk thickness increases\footnote{In contrast,
\citet{nob+2010} find that the stress profile is almost completely
independent of disk thickness.} \citep{kul+2011}, as anticipated in
previous work \citep{pac+2000,sha+2008b}.



\section{Uncertainties in Spin Estimates}\label{sec:error}

The bottom line of this section is that the error in $a_*$ is
dominated by the observational errors in the external input parameters
$D$, $i$ and $M$.  By comparison, the errors due to reliance on the NT
model, as well as on the disk atmosphere model that is used to correct
for the effects of spectral hardening, are less important.  Meanwhile,
the one significant question that hangs over most of the spin results
is the assumption that the black hole's spin vector is aligned with
the orbital angular momentum vector.  We discuss these points in turn.

\subsection{Observational Errors}\label{sec:error_obs}

In early work, the error in $a_*$ attributable to the uncertainties in
$D$, $i$ and $M$ was only crudely estimated
\citep{sha+2006,dav+2006,mcc+2006}.  However, in all subsequent work,
starting with \citet{liu+2008}, the error in the spin due to the
combined uncertainties in these three parameters has been computed in
detail via Monte Carlo simulations.  That the error budget for $a_*$
is dominated by the uncertainties in $D$, $i$ and $M$ has been
thoroughly demonstrated in recent work, which provides error estimates
for a very wide range of statistical and systematic errors associated
with (1) the details of the spectral models employed, (2) flux
calibration uncertainties, (3) the effects of a warm absorber, etc.
Instead of discussing such technical details here, we refer the reader
to Section~5 and Appendix~A in \citet{ste+2011} and Section 5 in
\citet{gou+2011}.

\subsection{Errors from the Novikov-Thorne Model}\label{sec:error_NT}

As described in Section~\ref{sec:cf_theory}, the NT model is robust
and makes very few untested assumptions. It is true that some
properties of the disk, e.g., the density and temperature of gas at
the disk mid-plane, depend on the magnitude of the viscosity parameter
$\alpha$, but the all-important luminosity profiles shown in
Figure~\ref{fig:RISCO}b do not. These profiles are a direct
consequence of energy conservation -- gas drifts inward, it converts
gravitational potential energy into orbital kinetic and gas thermal
energy, and the latter is radiated. This physics is independent of the
value of $\alpha$, or even the validity of the $\alpha$ prescription.

The NT model assumes that dissipated energy is radiated by the gas
locally at the same radius. This is a very safe assumption.  The
cooling time of the gas in a thin accretion disk is approximately
$(H/R)^2$ times the viscous radial advection time.  For the disks of
interest to the continuum-fitting method ($H/R < 0.1$), this means
that cooling is about 100 times faster than energy advection and hence
very local.

Another approximation in the NT model is the neglect of disk
self-irradiation. This is acceptable near the peak of the luminosity
profile, where local energy dissipation greatly exceeds
irradiation. However, it is less safe at larger radii.  The models
{\sc kerrbb} and {\sc kerrbb2} (Section~\ref{sec:cf_practice}) include
self-irradiation consistently.  In practice, self-irradiation seems to
have a minor effect on spin estimates.

As already discussed, the NT model assumes zero torque at the
ISCO. Although this approximation turns out to be less severe than one
might have anticipated (Figure~\ref{fig:RISCO}b), we still expect it
to have some effect on the continuum-fitting method. Several authors
\citep{kul+2011,nob+2011,zhu+2012} have investigated this issue
quantitatively. The general consensus is that the zero-torque
approximation introduces uncertainties in spin estimates of around
$\Delta a_*\sim0.1$ for low spin values $a_* < 0.5$ and much smaller
errors as $a_*\to1$.  For example, \citet{kul+2011} estimate for $a_*
=$ 0, 0.7, 0.9 and 0.98 that the respective values of $\Delta a_*$ are
0.11, 0.06, 0.014 and 0.007. (\citealt{nob+2011} estimate $\Delta a_*
\sim 0.2-0.3$ for $a_*=0$.) These results are for a disk inclination
angle of $i=60^{\circ}$ and a disk thickness of $H/R = 0.05$, which
corresponds to $L/L_{\rm Edd} \sim 0.35$ (see Table 1 in
\citealt{zhu+2012}).  The errors are more severe for thicker disks.
Meanwhile, the results quoted here are for the thinnest,
lowest-luminosity disks simulated to date; presently, it is not
practical to resolve the MRI turbulence in thinner disks.

Not only are these estimates of the NT model errors significantly less
than the observational errors presented in Section~\ref{sec:results},
they are overestimates of the model errors because the
continuum-fitting method is applied only to very thin disks: A strict
requirement of the method is $L/L_{\rm Edd} < 0.3$, while most
measurements are based on spectral data with $L/L_{\rm
Edd}~\simless~0.1$.  Looking to the future, it might be possible to do
better by replacing the NT model with a more accurate simulation-based
model \citep[e.g.,][]{pen+2012}, but this step is not presently
warranted.  In conclusion, all continuum-fitting spin
measurements published to date (see Section~\ref{sec:results}) are
based on the NT model which systematically overestimates the spin;
however, this source of error is presently small compared to the
observational errors. 

\subsection{Errors from the Disk Atmosphere Model}\label{sec:error_atmosph}

An essential cornerstone of the continuum-fitting method is a reliable
model of the disk's atmosphere.  Such a model is {\sc bhspec}
\citep{dav_hub+2006}, which can be used either alone or including the
effect of self-irradiation via {\sc kerrbb2}.  {\sc bhspec}, which is
quite sophisticated and includes a wide range of physical effects, is
based on the non-LTE radiative transfer code {\sc tlusty}
\citep{hub+1995}, which was originally developed for stellar
atmospheres.


At a given location on an accretion disk, {\sc bhspec} computes the
emitted spectrum using three supplied parameters: the effective
temperature $T_{\rm eff}$ defined such that radiative flux $F=\sigma
T_{\rm eff}^4$, local vertical gravity parameter $Q$, and disk column
density $\Sigma$. As discussed in previous sections, a robust estimate
of $T_{\rm eff}$ can be obtained from the NT model, while the
parameter $Q$ is calculated directly from the Kerr metric. The main
uncertainty is in the value of $\Sigma$.

In standard disk theory, $\Sigma$ varies inversely as the viscosity
parameter $\alpha$ and is thus quite uncertain.  Fortunately, in the
case of optically thick disks (which all thermal state disks are)
$\Sigma$ has only a weak effect on the emerging spectrum. This is
analogous to the case of a star where the spectrum depends on the
effective temperature and surface gravity, but not at all on the
optical depth to the stellar core, which is effectively infinite. The
optical depth through a disk is not quite infinite, hence there is
some spectral dependence on $\Sigma$.  However, this dependence is
weak for models with $L/L_{\rm Edd} < 0.3$
\citep{dav_hub+2006,don+2008}.

For the same reason, details of exactly how viscous heating is
distributed vertically within the disk are unimportant. So long as
energy dissipation occurs in the disk interior at optical depths
greater than a few, the emerging spectrum depends only on $T_{\rm
eff}$ and $Q$ \citep{dav+2005,dav+2009}. This is not true if there is
substantial energy dissipation close to or above the
photosphere. Disks in the thermal state probably do not have such
dissipation since their spectra show very little hard ``coronal''
emission \citep{rem+2006}. Whatever little coronal emission is present
is fitted for via a model for the Compton power law such as {\sc
simpl} (Section~\ref{sec:cf_practice}).

The standard {\sc bhspec} model assumes hydrostatic equilibrium and
does not include the force from magnetic fields.  However, numerical
simulations \citep[e.g.,][]{hir+2009} indicate that the photospheric
surface regions show modest deviations from hydrostatic equilibrium
and are primarily supported by magnetic forces.  Including these
effects in {\sc bhspec} generally leads to a modest ($< 10\%$)
increase in the spectral hardness \citep{dav+2009}.  The effects of
irradiation (both self- and from a corona), which have not yet been
rigorously explored, may also lead to a slight hardening of the
spectrum.
In summary, while there are uncertainties associated with the disk
spectral model used in the continuum-fitting method, it appears
unlikely that the resulting errors in $R_{\rm in}$ are more than 10\%,
which for low values of spin implies $\Delta{a_*}\sim0.1$, decreasing
as $a_*\to1$ (Figure~\ref{fig:RISCO}a).


\subsection{Assumption of Spin-Orbit Alignment}\label{sec:error_align}

In determining the spins of eight of the ten black holes (see
Section~\ref{sec:results}), it is assumed that the plane of the inner
X-ray-emitting portion of the disk is aligned with the binary orbital
plane, whose inclination angle $i$ is determined from optical
observations \citep[e.g.,][]{oro+2011a}.  However, if a black hole's
spin is misaligned with the orbital vector, this will warp a thin disk
because the Bardeen-Petterson effect will force the inner disk to
align with the black hole spin vector \citep{bar+1975}\footnote{While
thin disks are subject to warping, thick disks are not
\citep{dex+2011}.}.  An error in estimating the inclination of the
inner disk of $\sim10^{\circ}$ or more, resulting from an erroneous
use of $i$ as a proxy for the inclination of the inner disk, would
substantially corrupt most continuum-fitting measurements of
spin\footnote{Unfortunately the continuum-fitting method cannot fit
for the inclination of the inner disk because there is a degeneracy
between the inclination and spin parameter \citep{lil+2009}.}.

There is evidence for gross spin-orbit misalignment for one transient
system (SAX J1819.3-2525); however, this evidence is weak
\citep{nar+2005}.  For the transients generally, more recent evidence,
which is summarized in Section 1 of \citet{ste_mcc+2012}, argues in
favor of alignment.  Briefly, the timescale for accretion to torque
the black hole into alignment is estimated to be $\sim 10^6-10^8$
years, which is short compared to the typical lifetime of a transient
system (Section 2.2).  In the case of the persistent supergiant
systems, there is some evidence that their more massive black holes
are formed by direct, kickless collapse \citep{mir+2003,rei+2011}.
Finally, a population synthesis study based on a maximally
conservative (i.e., minimum-torque) assumption indicates that the spin
axes of most black hole primaries will be tilted less than
$10^{\circ}$ \citep{fra+2010}.

In determining the spins of the remaining two black holes (see
Section~\ref{sec:results}), the inclination of the inner disk is taken
to be the inclination $\theta$ of the radio or X-ray jet axis, which
is presumed to be aligned with the black hole's spin axis.  The jet
inclination angle for these microquasars, GRS~1915+105 and H1743--322,
was determined by modeling proper-motion data derived from radio and
X-ray observations \citep{mir+1994,fen+1999,ste+2012a}.  Fortunately,
radio/X-ray jet data have also yielded a strong constraint on $\theta$
for a third microquasar, XTE J1550--564, thereby providing a rare
opportunity to check directly the assumption of spin orbit alignment
because its orbital inclination angle $i$ has also been measured
\citep{oro+2011b}.  In this case, \citet{ste_mcc+2012} find no
evidence for misalignment and place an upper limit on the difference
between the spin and orbital inclinations of $|\theta - i| < 12$~deg
(90\% confidence).


\section{Results and Discussion}\label{sec:results}

Table~1 lists the masses and spins of ten stellar black holes.  By
virtue of the no-hair theorem, this table provides complete
descriptions of each of these ten black holes.  The spins span the
full range of prograde values, and the masses range from 6 to
16$~M_{\odot}$.  In addition to the continuum-fitting spin data in
Table~1, \citet{gie+2001} provide preliminary estimates for the spins
of LMC X-1 and GRO J1655--50, \citet{kol+2010} report a hard upper
limit of $a_*<0.9$ on the spin of GX~339--4, and \citet{now+2012}
argue that the spin of 4U~1957+11 is extreme.  Concerning 4U~1957+11,
it is unclear if the compact object is a black hole, and the key
parameters $D$ and $M$ are essentially unconstrained.  Finally,
\citet{mid+2006} find an apparently moderate value of spin for
GRS~1915+105, which is at odds with the extreme value in Table~1;
Middleton et al. obtained a depressed value of spin because they
relied on high-luminosity data, as explained in Section~5.3 in
\citet{mcc+2006}.

Caution is required in considering the errors for the values of spin
quoted in Table 1 assuming that they are Gaussian, particularly for
$a_*~\simgreat~0.7$.  Note in Figure~\ref{fig:RISCO}a how insensitive
$a_*$ is to large changes in the observable $R_{\rm ISCO}$ as $a_*$
approaches unity.  As a consequence of this limiting behavior of
$a_*$, doubling a 1$\sigma$ error to approximate a 2$\sigma$ error can
lead to nonsense.  For example, formally increasing the nominal spin
of LMC X-1 ($a_*=0.92$; Table 1) by doubling the 1$\sigma$ error
($\Delta a_* = 0.05$) implies a 2$\sigma$ upper limit of $a_*<1.02$,
whereas the correct 2$\sigma$ upper limit is $a_*<0.98$ (see Figure 8
in \citealt{gou+2009}).

\subsection{The Persistent Systems vs. the Transients}\label{sec:results_compare}

There is a dichotomy between the black holes in persistent systems and
those in transients, both in their masses and their spins (Table~1).
Considering spin first, the three persistent black holes all have high
or extreme spins.  In contrast, the spins of the transient black holes
range widely: Four have spins consistent with zero, two have
intermediate values of spin, and one is a near-extreme Kerr hole.  The
dichotomy is sharpened if one considers six additional transient black
holes all of whose spins are predicted to be $a_*~\simless~0.8$
\citep{ste+2013} based on a fitted correlation between radio power and
spin (Section~\ref{sec:jets_correlate}).

Not only are the persistent black holes rapidly spinning, they are
also massive -- $11-16~M_{\odot}$ -- compared to the transient black
holes.  The masses of the transients are significantly lower and,
remarkably, their mass distribution is narrow: $7.8\pm1.2~M_{\odot}$
\citep{oze+2010,far+2011}.

\begin{table*}
\caption{The masses and spins, measured via continuum-fitting, of ten stellar black holes$^a$.
\label{tab:results}}
\footnotesize
\begin{tabular}{lccl}
\hline
\noalign
{\vspace{1mm}}
System & $a_*$ & $M/M_{\odot}$ & References \\
\noalign
{\vspace{-2.5mm}}
&&& \\
\hline
\noalign
{\vspace{1.2mm}}
Persistent &&& \\
\noalign
{\vspace{1mm}}
\hline
\noalign
{\vspace{1.2mm}}
Cyg X-1 & $> 0.95$ & $14.8\pm1.0$ & \citealt{gou+2011}; \citealt{oro+2011a} \\
{\vspace{-3.5mm}}
&&& \\
\noalign
{\vspace{0.9mm}}
LMC X-1 & $0.92_{-0.07}^{+0.05}$ & $10.9\pm1.4$ & \citealt{gou+2009}; \citealt{oro+2009} \\
\noalign
{\vspace{0.9mm}}
M33 X-7 & $0.84\pm0.05$ & $15.65\pm1.45$ & \citealt{liu+2008}; \citealt{oro+2007} \\
\noalign
{\vspace{1mm}}
\hline
\noalign
{\vspace{1mm}}
Transient &&& \\
\noalign
{\vspace{1mm}}
\hline
\noalign
{\vspace{1.2mm}}
GRS 1915+105 & $> 0.95^b$ & $10.1\pm0.6$ & \citealt{mcc+2006}; \citealt{stg+2013} \\
{\vspace{-3.5mm}}
&&& \\
\noalign
{\vspace{0.9mm}}
4U 1543--47 & $0.80\pm0.10^b$ & $9.4\pm1.0$ & \citealt{sha+2006}; \citealt{oro+2003} \\
\noalign
{\vspace{0.9mm}}
GRO J1655--40 & $0.70\pm0.10^b$ & $6.3\pm0.5$ & \citealt{sha+2006}; \citealt{gre+2001} \\
\noalign
{\vspace{0.9mm}}
XTE J1550--564 & $0.34_{-0.28}^{+0.20}$ & $9.1\pm0.6$ & \citealt{ste+2011}; \citealt{oro+2011b} \\
\noalign
{\vspace{0.9mm}}
H1743--322 & $0.2\pm0.3$ & $\sim8^c$ & \citealt{ste+2012a} \\
\noalign
{\vspace{0.9mm}}
LMC X-3 & $< 0.3^d$ & $7.6\pm1.6$ & \citealt{dav+2006}; \citealt{oro+2003} \\
\noalign
{\vspace{0.9mm}}
A0620--00 & $0.12\pm0.19$ & $6.6\pm0.25$ & \citealt{gou+2010}; \citealt{can+2010} \\ 
\noalign
{\vspace{0.7mm}}
\hline
\end{tabular}
\vskip 2mm
Notes: \newline
$^a$~Errors are quoted at the 68\% level of confidence, except for the
three spin limits, which are estimated to be at the 99.7\% level of confidence. \newline
$^b$~Uncertainties greater than those in papers cited because early error estimates were crude. \newline
$^c$~Mass estimated using an empirical mass distribution \citep{oze+2010}. \newline
$^d$~Preliminary result pending improved measurements of $M$ and $i$.
\end{table*}

\subsection{Prograde Spins that obey the Kerr Bound}\label{sec:results_prograde}

The lack of negative spins in Table~1 may be the result, in a close
binary system, of the expected alignment of the spin of the black hole
progenitor with the orbital angular momentum, and it may also indicate
that black hole kicks are not strong enough to flip the black hole
into a retrograde configuration.  While interesting that there are no
negative spins, it is equally interesting that the spins of all ten
black holes obey the Kerr bound $|a_*| < 1$.  In particular, if the
distances to either Cyg X-1 or GRS~1915+105 were $\sim30$\% less than
the best current estimates, then it would be impossible to fit the
data with the {\sc kerrbb2} model, which only accommodates spin values
$a_*<1$.  Because the observed values of each of the three external
fit parameters ($D$, $i$ and $M$) place hard constraints when fitting
the data, a failure to fit a spectrum that requires $a_*>1$ has the
potential to falsify the spin model.  For a discussion of this point
in relation to the near-extreme Kerr hole GRS~1915+105, see
Section~6.4 in \citet{mcc+2006}.

\subsection{The High Natal Spins of the Persistent Black Holes}\label{sec:results_natal}

It is reasonable to conclude that the black holes in the persistent
systems were born with high spins because their host systems are too
young for these black holes to have been spun up by accretion torques.
Consider, for example, the persistent system Cyg~X-1 \citep{gou+2011}:
For its black hole to achieve its present spin of $a_*>0.95$ via disk
accretion, an initially nonspinning black hole would have had to
accrete $>7.3~M_{\odot}$ from its donor \citep{bar+1970,kin+1999} to
become the $14.8~M_{\odot}$ black hole we observe today.  However,
even at the maximum (Eddington-limited) accretion rate this would
require $>31$ million years, while the age of the system is between
4.8 and 7.6 million years \citep{won+2012}.  Likewise for M33 X-7 and
LMC X-1, the corresponding minimum spin-up timescales are $>17$ and
$>25$ million years, respectively, while the respective ages of the
systems are $\simless~3$ and $\simless~5$ million years
\citep{gou+2011}.  It therefore appears that the spins of these
systems must be chiefly natal, although possibly such high spins could
be achieved during a short-lived evolutionary phase of hypercritical
accretion \citep{mor+2008}.

\subsection{Applications}\label{sec:results_apply}

The data in Table~1 have a number of applications to physics and
astrophysics, both immediate and potential.  In physics, a high goal
is to use such data as a springboard to test the no-hair theorem (see
Section~\ref{sec:conclusion}), and the foundation for any such test is
high-quality measurements of mass and spin for a good sample of black
holes.  In astrophysics, knowledge of the spins of stellar black holes
is crucial for example in constraining models of gamma-ray burst
sources \citep{woo+1993,mac+1999,woo+2006}; supernovae and black hole
formation \citep{lee+2002,won+2012}; exotic black hole states and
state transitions \citep{rem+2006}; and in informing LIGO/VIRGO
modelers who are computing gravitational-wave signals
\citep{cam+2006}.  A central question, which we turn to in the next
section, is the role of spin in powering jets.


\section{Jet Power and Black Hole Spin}\label{sec:jets}

Since the spin parameter $a_*$ is one of only two numbers that
completely characterize a black hole (mass $M$ being the other), it
stands to reason that it should influence at least some observational
properties of the hole. The most widely discussed connection is to
relativistic jets.

The story goes back to \citet{pen+1969} who showed that a spinning
black hole has free energy that can in principle be tapped by
specially prepared infalling particles. Although Penrose's specific
proposal is not considered promising, the idea of extracting energy
from spinning black holes has stuck and has become popular in
astrophysics.  \citet[][see also \citealt{dam+1978}]{ruf_wil1975} and
\citet{bla+1977} suggested a specific mechanism whereby a force-free
poloidal magnetic field around a spinning black hole is twisted by
frame dragging, thereby producing outgoing Poynting flux along twin
jets. We refer to this as the generalized Penrose process.

GRMHD simulations of accreting black holes have found MHD jets forming
spontaneously from generic initial conditions (e.g.,
\citealt{koi+2002, mck_gam2004, mck+2005, bec+2008,
mck_bla2009}). Moreover, in one particular simulation involving a
rapidly spinning black hole and a strong poloidal field,
\citet{tch+2011} showed that the power carried by the jet exceeded the
total rest mass energy of accreted gas, meaning that the jet extracted
energy from the spinning black hole.

On the observational front, until recently there was no empirical
evidence for a connection between black hole spin and relativistic
jets. We discuss here the first such evidence.

\subsection{Two Kinds of Jets in Black Hole Binaries}\label{sec:jets_2kinds}
\label{twojets}

\citet{fen+2004} identified a number of systematic properties in the
radio emission of black hole binary jets.  They showed that there are
two kinds of jets, which we refer to as ``steady jets'' and
``ballistic jets,'' each associated with a specific spectral state of
the X-ray source. Although we discuss both kinds of jets for
completeness, our focus here is on the ballistic jet.  

The steady jet is observed as a continuous outflow of plasma in the hard
spectral state \citep[for a discussion of spectral states in black
hole binaries, see][]{rem+2006}.  This jet is small-scale, being
observable only out to a few tens of AU, and it appears not to be very
relativistic.  It is present at the very start of a transient's
outburst cycle.  Referring to the X-ray light curves for XTE~J1859+226
in Figure~\ref{fig:j1859}a, the jet is present during the first few
days when the hard flux (panel a) is most intense and the 2--12 keV
flux (panel b) is increasing rapidly.  It then disappears, and it
returns only near the end of the outburst cycle (beyond the right edge
of the plot).  The steady jet is seen in all transients at low values
of $\dot{M}$.

The far more dramatic ballistic jet is launched when a transient goes
into outburst \citep{fen+2004}.  This powerful transient jet usually
appears near (or soon after) the time of outburst maximum, as the
source switches from its initial hard state to a soft state via the
``steep power-law'' state.  Ballistic jets manifest themselves as
blobs of radio (and occasionally X-ray) emitting plasma that move
ballistically outward at relativistic speeds (Lorentz factor
$\Gamma>2$).  They are often observed out to distances of order a
parsec.  Because ballistic jets resemble the kpc-scale jets seen in
quasars, black hole binaries that produce them are called microquasars
\citep{mir+1999}.

Ballistic jet ejection occurs at a very specific stage during the
spectral evolution of a given system \citep{fen+2004}.  In
Figure~\ref{fig:j1859}, the strong spike in the radio light curve
(panel d), which is characteristically delayed relative to the
corresponding spike in the X-ray luminosity (panel b), is associated
with a ballistic jet.  As most clearly demonstrated for the
prototypical microquasar GRS 1915+105 \citep{fen_bel+2004}, this
ejection stage appears to correspond to the inward-moving inner edge
of the accretion disk reaching the ISCO, which apparently results in
some violent event that launches a large-scale relativistic blob.  

On general principles, one expects jet power to depend on a black
hole's mass $M$ and spin $a_*$, and the mass accretion rate $\dot{M}$,
plus other factors such as the strength and topology of the magnetic
field. If one wishes to investigate the dependence of jet power on
$a_*$, one needs first to eliminate the other variables.

For steady jets, $\dot{M}$ spans a wide range, and it is not
straightforward to eliminate the effects of this variable.  It is
possible to do this, however, by using the disk X-ray luminosity as a
proxy for $\dot{M}$ (e.g., \citealt{hei_sun2003, mer+2003, fal+2004,
fen+2004, fen_bel+2004}), but one must have knowledge of the radiative
efficiency of the disk, which is generally both low and variable in
the hard state where the steady jet forms \citep[see][for a discussion
of radiatively inefficient accretion in the hard state]{nar+2008}.
The procedure is relatively uncertain and it is difficult to obtain
robust results.  Nevertheless, \citet{fen+2004} have performed such a
study and have claimed that there is no evidence in the data that the
power of a steady jet depends on spin.

Ballistic jets on the other hand invariably occur near the peak of
transient outbursts. \citet{ste+2013} have shown that during major
outbursts the peak disk luminosities in various transients are near
the Eddington limit and are clustered within a factor of $\sim2$ in
luminosity, which means that these systems behave for all purposes
like ``standard candles.''  This crucially allows one to compare the
power of ballistic jets observed for different black holes at the same
$\dot{M}$, namely $\dot{M}~{\sim}~\dot{M}_{\rm Edd}$. In addition, all
black holes in transients have similar masses to better than a factor
of two \citep{oze+2010}; furthermore, it is easy to correct for any
mass dependence (see below).  This leaves $a_*$ (with magnetic field
as a wild card) as the sole remaining parameter that could have any
influence on jet power.

Let us define the jet efficiency factor $\eta$ of a ballistic jet,
\begin{equation}
\eta_{\rm jet}(a_*) = \langle L_{\rm jet}\rangle/\langle\dot{M}\rangle c^2,
\end{equation}
where $\langle L_{\rm jet}\rangle$ is the time-average kinetic
luminosity flowing out through the jet and $\langle\dot{M}\rangle c^2$
is the time-average rate at which rest-mass energy flows into the
black hole.  Using (1) radio luminosity as a proxy for $L_{\rm jet}$
\citep{nar+2012} and (2) observed values for the peak radio
luminosities for five black holes that are all accreting at
${\sim}\dot{M}_{\rm Edd}$ \citep{ste+2013}, one can infer directly how
jet luminosity depends on spin, as we discuss in the following section.

\subsection{Correlation Between Spin and Ballistic Jet Power}\label{sec:jets_correlate}

A typical ballistic jet blob is initially optically thick and has a
low radio power.  As the blob moves out and expands, the larger
surface area causes its radio power to increase. This continues until
the blob becomes optically thin, after which the flux declines
rapidly.  The overall behavior is generally consistent with an
expanding conical jet \citep{van+1966,hje+1988}.  Moreover, as
discussed in Section~\ref{subsec:synch}, the peak radio
luminosity is expected to scale more or less linearly with the jet
kinetic energy or kinetic luminosity.  Thus, peak radio luminosity is
a good proxy for jet kinetic luminosity.

\citet{nar+2012} considered the peak radio luminosities of ballistic
jet blobs in four transients, A0620--00, XTE J1550--564, GRO
J1655--40, GRS 1915+105, and showed that they correlated well with the
corresponding black hole spins measured via the continuum-fitting
method\footnote{In the case of a fifth transient, 4U1543--47, radio
observations did not include the peak of the light curve, so one could
only deduce a lower limit to the jet power. Note that the radio peak
can be very narrow in time, e.g., $\approx~1$-day in the case of
XTE~J1859+226 (Figure~\ref{fig:j1859}), so one requires dense radio
monitoring to catch the peak.}.  Later, \citet{ste+2013} included a
fifth transient, H1743--322, whose spin had been just measured.
Figure~\ref{fig:jet}a shows a plot of the black hole spins of these
five objects versus a measured quantity called ``Jet Power,'' which
refers to the radio luminosity $\nu L_\nu = (\nu S_\nu)D^2/M$ (here,
not corrected for beaming), where $\nu=5$\,GHz is the radio frequency,
$S_\nu$ is the flux density in Jy at the peak of the ballistic-jet
radio light curve, $D$ is the distance in kpc, and $M$ is the black
hole mass in solar units\footnote{The scaling by mass is sensible
because the sources are near the Eddington luminosity limit, which is
proportional to mass.  However, since the masses of the black holes
differ little (Table 1), the results would be virtually identical if
the mass scaling were eliminated.}. That is, the proxy adopted for jet
kinetic luminosity is simply the peak radio luminosity at
5~GHz\footnote{None of the results change if one chooses a different
reference frequency, e.g., 1.4~GHz or 15~GHz.}.  Figure~\ref{fig:jet}a
shows unmistakable evidence for a strong correlation between Jet Power
and $a_*$. Note that Jet Power varies by nearly three orders of
magnitude as the spin parameter varies from $\approx0.1-1$.

\begin{figure}
  \includegraphics[width=0.45\textwidth]{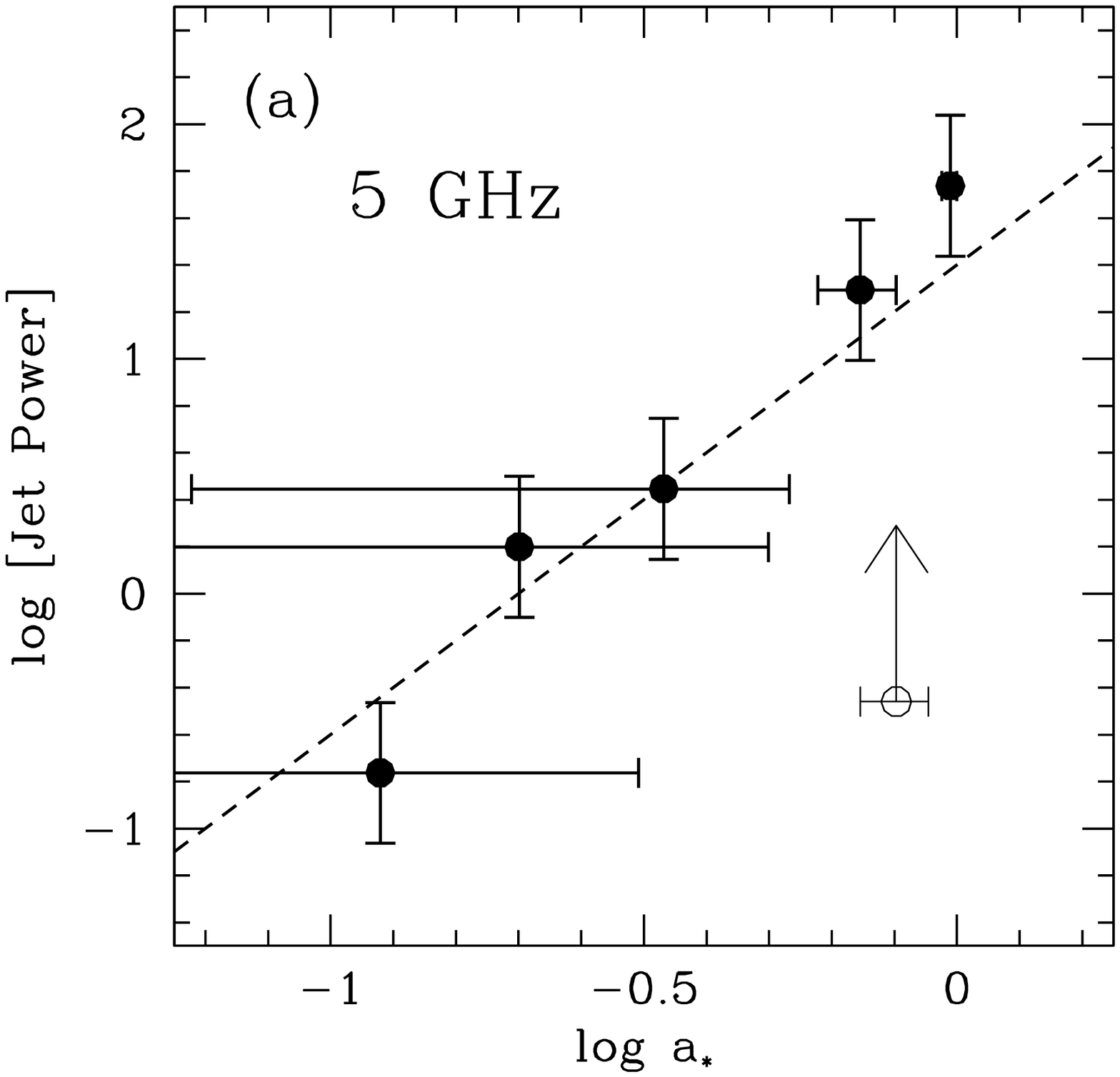}
  \includegraphics[angle=90,width=0.55\textwidth]{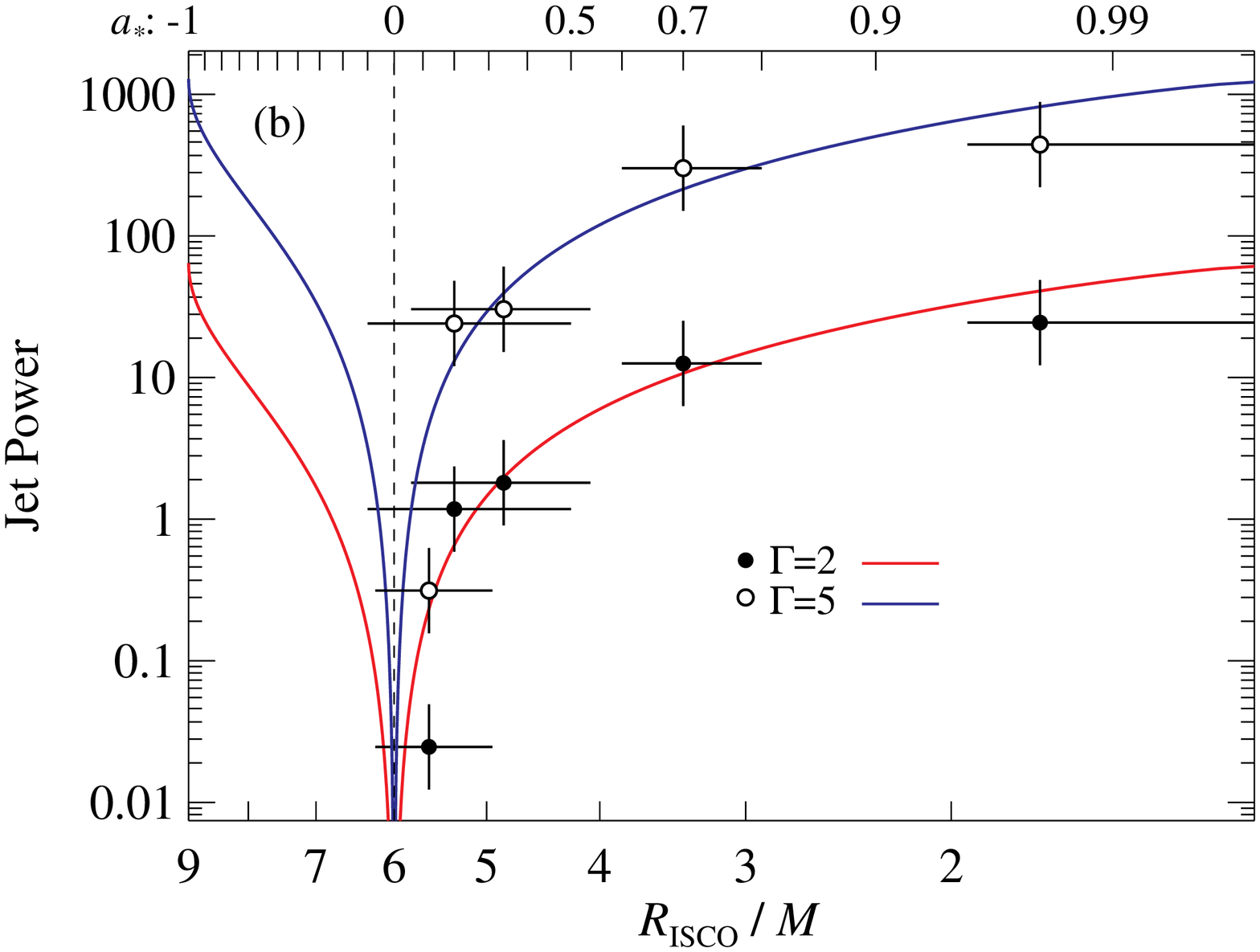}
\caption{(a) Plot of the quantity Jet Power, which measures the 5~GHz
  radio luminosity at light curve maximum, versus black hole spin,
  measured via the continuum-fitting method for five transients
  \citep{nar+2012, ste+2013}. The dashed line has slope equal to 2.
  (b) Plot of Jet Power versus $R_{\rm ISCO}/(GM/c^2)$.  Here the
  radio luminosity has been corrected for beaming assuming a bulk
  Lorentz factor $\Gamma=2$ (filled circles) or $\Gamma=5$ (open
  circles). The solid lines correspond to Jet Power $\propto
  \Omega_{\rm H}^2$, where $\Omega_{\rm H}$ is the angular frequency
  of the horizon \citep{ste+2013}.}
\label{fig:jet}
\end{figure}

The uncertainty in the estimated values of Jet Power, which is
difficult to assess, is arbitrarily and uniformly assumed to be a
factor of two \citep{nar+2012}.  The very unequal horizontal error
bars in Figure~\ref{fig:jet}a are a feature of the continuum-fitting
method of measuring $a_*$. Recall that the method in effect measures
$R_{\rm ISCO}$ and then deduces the value of $a_*$ using the mapping
shown in Figure~\ref{fig:RISCO}a.  Since the mapping is highly
non-linear, especially as $a_*\to1$, comparable errors in $R_{\rm
ISCO}$ correspond to vastly different uncertainties in $a_*$.  In
addition, the use of log $a_*$ along the horizontal axis tends to
stretch error bars excessively for low spin values.  This point is
clarified by considering Figure~\ref{fig:jet}b, based on
\citet{ste+2013}.  Here the horizontal axis tracks $\log R_{\rm ISCO}$
rather than $\log a_*$, and the horizontal error bars are therefore
more nearly equal. The key point is, regardless of how one plots the
data, the correlation between Jet Power and black hole spin appears to
be strong.

\subsection{What Does it Mean?}\label{sec:jets_mean}

Assuming the correlation shown in Figure~\ref{fig:jet} is real, there
are two immediate implications: (i) Ballistic jets in black hole
binaries are highly sensitive to the spins of their underlying black
holes, presumably because these jets derive their power directly from
the spin energy of the hole \citep[\'a la][]{pen+1969, ruf_wil1975,
bla+1977}.  (ii) Spin estimates of stellar black holes obtained
via the continuum-fitting method are sufficiently reliable to reveal
this long-sought connection between relativistic jets and black hole
spin.

With respect to item (i), we note that the mere existence of a
correlation does not necessarily imply that the Penrose process is at
work. We know that the gravitational potential in the inner disk
deepens with increasing black hole spin, since the inner radius of the
disk ($R_{\rm ISCO}$) shrinks with increasing $a_*$
(Figure~\ref{fig:RISCO}a).

Could this disk-related effect be the reason for the increasing jet
power? It seems unlikely. The radiative efficiency of a Novikov-Thorne
thin accretion disk increases only modestly with spin; for the five
objects shown in Figure~\ref{fig:jet}, the radiative efficiencies are
0.061, 0.069, 0.072, 0.10 and 0.19, respectively, varying by a factor
of $\approx3$.  It seems implausible that a disk-powered jet could
vary in radio Jet Power by the orders of magnitude seen in
Figure~\ref{fig:jet}.

In contrast, any mechanism that taps directly into the black hole spin
energy via something like the Penrose process can easily account for
the observed large variation in Jet Power. For instance, the
generalized Penrose effect predicts that the jet efficiency factor
should vary as $\eta_{\rm jet}\propto a_*^2$
\citep{ruf_wil1975,bla+1977} or, more precisely, as $\eta_{\rm
jet}\propto \Omega_{\rm H}^2$ \citep{tch+2010}, where $\Omega_{\rm H}$
is the angular frequency of the black hole horizon\footnote{The two
scalings agree for small values of $a_*$, but differ as $a_*\to1$.},
\begin{equation}
\Omega_{\rm H} = \frac{c^3}{2GM}\,\left(\frac{a_*}{1+\sqrt{1-a_*^2}}\right).
\end{equation} 
The dashed line in Figure~\ref{fig:jet}a corresponds to Jet Power
$\propto a_*^2$ and the solid lines in Figure~\ref{fig:jet}b to $\propto
\Omega_{\rm H}^2$.  The observational data agree remarkably well with
these scalings, strongly suggesting that a Penrose-like process is in
operation.

While the above conclusions are highly satisfying, one should not
discount other possibilities. First, the correlation shown in
Figure~\ref{fig:jet} is based on only five objects. Although this is
mitigated by the very wide range of Jet Powers, the correlation might
become weaker with the next spin measurement.  Second, the correlation
might arise if Jet Power and spin are each correlated with some third
parameter.  For instance, it is intriguing that the binary orbital
periods of the five transients under consideration increase
systematically with Jet Power.  One could imagine scenarios in which
the energy of ejected blobs depends on the size of the accretion disk,
which depends in turn on the orbital period. However, it is less easy
to see why the values of $a_*$ measured with the continuum-fitting
method should correlate with the orbital period, while conspiring to
produce the scaling predicted by \citet{bla+1977}.

Returning to the subject of steady jets, it is interesting to consider
why there is apparently no correlation between jet radio luminosity
and spin \citep{fen+2010,rus+2013}.  One likely answer is that steady
jets and ballistic jets are produced via very different
mechanisms. Perhaps ballistic jets are launched within a few
gravitational radii, near $R_{\rm ISCO}$, where the black hole spin
could plausibly have a strong effect, whereas steady jets in the hard
state originate much further out at radii $\sim 10-100$~GM/c$^2$
\citep[e.g.,][]{mar+2005}, where the effects of spin are relatively
weak. In support of this explanation, ballistic jets are definitely
relativistic, with Lorentz factors of up to several
\citep{fen+2004,fen+2006}, whereas there is little evidence that
steady jets are relativistic \citep{gal+2003,nar+2005}.

\subsection{A Challenge}\label{sec:jets_challenge}

Based on an analysis of a heterogeneous data sample of uneven quality,
\citet{fen+2010} claim that there is no evidence for a correlation
between the power of ballistic jets and black hole spin.  The
substantial difference between the results obtained by these authors
and \cite{nar+2012} is, in the end, determined by the quantity used to
represent jet power (which is discussed further in
Section~\ref{subsec:synch}). Fender et.\ al compute jet power from the
peak radio luminosity and rise time of a particular synchrotron event;
adopt a formula relating jet power to X-ray luminosity, namely,
log$_{\rm 10}L_{\rm jet} = c~+~0.5($log$_{\rm 10}L_{\rm x}~-~34)$; and
use the normalization constant $c$ as their proxy for the jet power.
Narayan and McClintock, on the other hand, use the model-independent
proxy discussed above, namely, the maximum observed radio luminosity
at 5~GHz.

Very recently, the three authors of \citet{fen+2010} have written a
second paper \citep{rus+2013} repeating their claim that there is no
evidence for a correlation between the power of ballistic jets and
black hole spin.  Therein, they challenge the methodologies and
findings of \cite{nar+2012} and \cite{ste+2013}.  A response to this
challenge is being readied (J.\ Steiner et al., in preparation);
meantime, the following is a preliminary sketch of the elements of
this response.  The following comments pertain only to ballistic jets
and continuum-fitting spin measurements. \citep[][also discuss steady
jets and Fe-line spin measurements]{rus+2013}.

\subsubsection{Significance of the Result}

\citet{rus+2013} contend that the empirical correlation shown in
Figure~\ref{fig:jet}a is only marginally statistically significant
($\approx90$\% confidence).  Their analysis is based on a Bayesian
linear regression model \citep{kel+2007}, which incorporates an
additional parameter that allows for intrinsic noise; this
putative and undefined source of noise is distinct from
measurement error.  Adopting precisely the procedures of Russell et
al., but using the \citet{ste+2013} data set, we confirm the
correlation at $\approx 95\%$ confidence
\footnote{The lower level of confidence reported by \citet{rus+2013}
is largely attributable to their use of radio data for a flare of
H1743--322 that is unrelated to the ballistic jet, namely a flare
event that occurred 28 days before the X-ray flux peaked
\citep{mcc+2009}.  We focus solely on post-Eddington radio flares.
The relevant radio flare event, which we consider, occurred 2.6 days
after X-ray maximum \citep{ste+2013}.}.~~In contrast, a traditional
analysis (i.e., a linear fit using two-dimensional error bars)
produces a correlation at $\approx 99.9\%$ confidence.  A comparably
strong result is achieved using the \citet{kel+2007} model if one
applies the usual Jeffreys noninformative prior to the intrinsic noise
term rather than the default flat prior, which preferentially selects
solutions corresponding to large estimates of intrinsic noise.

Of greater importance, \citet{rus+2013} consider only the empirical
correlation shown in Figure~\ref{fig:jet}a, and they disregard the
physical model which takes into account beaming effects, namely, the
model shown in Figure~\ref{fig:jet}b.  The application of this
physical model is based on four simple assumptions: (i) spin-orbit
alignment \citep{ste_mcc+2012}; (ii) $\Gamma$ is the same for all five
sources and bracketed between 2 and 5 \citep{fen_lwnbk+2006}; (iii)
jet power is proportional to $\Omega_{\rm H}^2$
\citep{bla+1977,tch+2010}; and (iv) radio luminosity can be used as a
proxy for jet power (Sections~\ref{sec:jets_correlate} and
\ref{subsec:synch}).  Fitting the \citet{ste+2013} data set to this
model, with the model normalization as its {\em sole} fit parameter,
one obtains good fits (shown in Figure~\ref{fig:jet}b) with
$\chi^2/\nu = 0.3$ and 0.5 for $\Gamma=2$ and 5, respectively.  This
relationship is determined over a span of $\sim 3$ orders of magnitude
in jet power, and over the full allowed range of prograde spins.  It
is reasonable to view this result as evidence that the
Blandford-Znajek model successfully describes the behavior of a
ballistic jet produced by a black hole transient as it approaches its
Eddington limit.

\subsubsection{Issue of Data Selection}

\citet{rus+2013} furthermore argue that \citet{nar+2012} and
\citet{ste+2013} omit data for several systems that should be included
in the correlation plots shown in Figure~\ref{fig:jet}.  These
additional data, which are plotted in Figure~1c in \citet{rus+2013},
destroy the clean correlation shown in Figure~\ref{fig:jet}a.
However, it is inappropriate to include these data, and we reject them
for the following reasons.

{\bf Cygnus X-1} radiates persistently at a few percent of Eddington
and, during its periods of jet ejection \citep{fen+2006}, its mass
accretion rate is both very low and poorly constrained.  At the same
time, as stressed in Section~\ref{sec:jets_2kinds}, comparing sources
at the same mass accretion rate, namely $\dot{M}~{\sim}~\dot{M}_{\rm
Edd}$, is the essential methodological requirement that eliminates the
otherwise unknown dependency of the jet efficiency $\eta_{\rm jet}$ on
$\dot{M}$ (see equation~1).  In order to include Cyg X-1 in the
\citet{ste+2013} sample, it would be necessary to know the precise
scaling of jet power with $\dot{M}$ and then to estimate $\dot{M}$ at
the time of jet ejection; $\dot{M}$ is $\sim1-2$ orders of magnitude
below Eddington and highly uncertain.  \citet{rus+2013} ignore this
problem.

For {\bf GRS~1124--68} and {\bf GS~2000+25}, \citet{rus+2013} adopt
continuum-fitting spin data that are completely unreliable.  These
data are from a pioneering, proof-of-concept paper on continuum
fitting \citep{zha+1997} whose authors note that the crucial ``system
parameters [i.e., $D$, $i$ and $M$] are mostly unknown.''  More
importantly, the values of spin adopted by Russell et al. for these
two sources were simply inferred by Zhang et al.\ from crude estimates
of the inner-disk radius $R_{\rm in}$ taken from a review paper
\citep{tan_lew+1995}; these estimates of $R_{\rm in}$ were, moreover,
computed using the nonrelativistic disk model {\sc diskbb} that
assumes a grossly incorrect inner-boundary condition \citep{zim+2005},
while neglecting the effects of spectral hardening.

{\bf GRS~J1655--40} had major outbursts in both 1994 and 2005.
\citet{rus+2013} plot in their Figure~1c a data point for the 2005
outburst which corresponds to a very faint radio flare.
\citet{nar+2012} and \citet{ste+2013} do not include data for the 2005
outburst in their sample because the radio coverage was too sparse:
The proximate observations that bracket the peak 4.3 Crab flare
\cite[see {\rm RXTE}/ASM plot in Figure~1 in][]{bro+2006} occur 2.3
days before and 4.7 days after the 2-12 keV maximum \citep[see
observation log in the NRAO VLA archive and web link to a plot of the
radio data in][]{rup+2005}.  We argue that the source produced a
bright and brief radio flare (e.g., like the 1-day radio spike
observed for XTE~J1859+226 shown in Figure~\ref{fig:j1859}d), which
was not observed because of the week-long gap in radio coverage.  For
the 2002 outburst of 4U~1543--47, we likewise hold that the radio
coverage was inadequate and the radio flare was missed, as discussed
in \citet{nar+2012}.

\subsubsection{Synchrotron Bubble Model}\label{subsec:synch}

Another point of contention is how to relate the radio luminosity $\nu
L_\nu$ observed at the peak of the light curve --- this is the
quantity ``Jet Power'' in Figure~6 --- to physical conditions in the
jet. \citet{ste+2013} used synchrotron theory with the following
standard assumptions: (1) The nonthermal radio-emitting electrons in
the jet blob have an energy distribution $N(\gamma)\sim \gamma^{-p}$
where $\gamma$ is the electron Lorentz factor in the blob frame; (2)
$p=5/2$ to be consistent with the synchrotron spectrum; (3) the
magnetic energy in the blob is in equipartition with the energy in the
nonthermal electrons; (4) there is one proton for each nonthermal
electron and the total energy of the blob is dominated by the kinetic
energy $E$ of the protons; and (5) at light curve maximum, the jet
blob transitions from optically thick to thin
\citep{van+1966}. Steiner et al.\ showed that $E \sim (\nu
L_\nu)^{1.2}$, i.e., the blob energy varies approximately linearly
with $\nu L_\nu$. They thus argued that the latter is a good proxy for
the former.

Why consider blob energy $E$? In the synchrotron bubble model
\citep{van+1966}, the jet ejection is some brief episode that is not
observationally resolved. Hence the total ejected energy is all that
one can measure.  \citet{rus+2013} (and references therein) focus on
the jet kinetic luminosity, $L_{\rm jet} =dE/dt \sim E/t_{\rm jet}$,
where $t_{\rm jet}$ is the duration of the jet ejection.  They further
assume that the jet is ejected continuously, with a constant
luminosity, until light curve maximum\footnote{The ejection apparently
shuts off, coincidentally, as the jet becomes optically thin.}. How
does $L_{\rm jet}$ depend on $\nu L_\nu$?  \citet{ste+2013} find that
the radius of the jet blob scales as $R \sim (\nu L_\nu)^{0.5}$.
Therefore, if the blob expands with some constant speed, say $c$, then
$L_{\rm jet} \sim E/(R/c) \sim (\nu L_\nu)^{0.7}$, i.e., jet kinetic
luminosity varies as $\nu L_\nu$ to a power somewhat less than
unity\footnote{Using somewhat different assumptions, \citep{wil+1999}
estimated that kinetic jet kinetic luminosities of radio galaxies
should vary as $(\nu L_\nu)^{6/7}$, i.e., a slope again close to but
less than unity.}.  The truth is probably somewhere in between this
result and that obtained for the blob energy $E$, i.e., very close to
a linear dependence. Note that \citet{rus+2013} ignore altogether the
fact that the jet blob transitions from optically thick to thin at
light curve maximum, thereby missing a key piece of information. As a
result, they do not have an analytic estimate of $R$ and need to
estimate $t_{\rm jet}$ from the poorly constrained rise time to
maximum of the radio flux.

\subsubsection{On Resolving the Controversy}

In our view, the significant challenge posed by Russell et al. (2013)
is whether GRO J1655-40 did or did not produce a strong radio flare
during its 2005 outburst (Section 7.4.2); a similar challenge is posed
for the 2002 outburst of 4U 1543-47 \citep{nar+2012}.  We maintain
that both sources produced bright radio flares, but that they were
missed because of the sparse radio coverage.  On the other hand,
\citet{rus+2013} and D. Russell (private communication) contend that
the radio coverage was adequate to detect the strong flares during
their decay phase, and they conclude that neither source produced a
strong flare.  This controversy cannot be firmly decided because the
radio data collected for these events are inadequate.

Fortunately, we can expect the controversy to be settled relatively
soon via radio observations of transient outburst events using new
facilities such as the MeerKAT array (assuming a continuing capability
to monitor the X-ray sky), and by the continual progress in obtaining
secure measurements of the spins of transient black holes.  MeerKAT, a
forerunner of the SKA, is an array of 64 dishes scheduled for
commissioning in 2014--2015 that will have outstanding sensitivity.
R.\ Fender and P.\ Woudt, the PIs of the science project ThunderKAT,
will obtain definitive measurements of all bright black-hole
transients at high cadence (R. Fender, private communication).



\section{Conclusions and Future Prospects}\label{sec:conclusion}

The continuum-fitting method has a number of virtues.  A principal one
is the simple and elegant model upon which it is based, namely the
model of a thin, viscous accretion disk.  This model was anticipated
and developed well before the existence of black holes was established
\citep[e.g.,][]{lyn+1969}.  Shortly thereafter, the analytic theory of
thin disks was fully developed, an effort that culminated in the
workhorse NT model \citep{nov+1973}.  The most important predictions
of the model have been validated recently via GRMHD simulations
(Sections~\ref{sec:isco_theory}~and~\ref{sec:error_NT}). This
venerable model, with the addition of a model of the disk's atmosphere
(Sections~\ref{sec:cf_theory}~and~\ref{sec:error_atmosph}),
measurements of three key parameters ($D$, $i$ and $M$), and suitable
X-ray data, allows one to estimate the inner-disk radius.  Meanwhile,
an abundance of strong observational and theoretical evidence allows
one to identify the inner-disk radius with the radius of the ISCO,
which is simply related to the spin of the black hole.

Another key virtue of the continuum-fitting method is an abundance of
data for which the thermal disk component is strongly dominant.  For
most stellar black holes, one has many suitable archival spectra for
which only a few percent or less of the thermal seed photons are
upscattered by a hot corona into a Compton tail component of emission.
In short: The continuum-fitting model is tried and true, and there is
an abundance of suitable data for many stellar black holes obtained
for the simplest and best understood state of an accreting black hole,
namely, an optically-thick thermal disk.

The spins and masses of ten stellar black holes are given in Table~1.
Their spins span the full range of prograde values, and their masses
range from $6-16~M_{\odot}$.  Setting aside the extreme spin of the
transient GRS~1915+105, the three persistent black holes have higher
spins and larger masses than their transient cousins.  Furthermore,
the high spins of these persistent, young black holes are unlikely to
have been achieved via accretion torques, which implies that their
spins are natal.  The spins and masses of the ten black holes in
Table~1 provide their complete description and are the essential data
for testing astrophysical models of how an accreting black hole
interacts with its environment.  They are likewise essential data for
underpinning the physical theory of black holes, and for ultimately
attempting tests of the no-hair theorem by, e.g., observing deviations
from the multipoles predicted by the Kerr metric, all of which are
functions of just the two parameters $a_*$ and $M$
\citep{vig+2010,joh+2010,joh+2011,bam+2013}.

To date, the most important application of the data in Table~1 is to
one class of jets, namely, ballistic jets that are produced in major,
Eddington-limited outbursts of black holes in transient systems.  For
such outbursts the peak radio luminosities of five of these
microquasars correlate strongly with their spins, increasing by a
factor of $\sim1000$ as spin increases from $\sim0$ to $>0.95$.
Meanwhile, a simple synchrotron jet model shows radio luminosity to be
a good proxy for jet power.  As \citet{nar+2013} discuss in detail,
the fitted relationship between jet power and spin
(Figure~\ref{fig:jet}b) is not only a validation of the classic model
of \citet{bla+1977}, it was also anticipated by GRMHD simulations
showing that a jet can extract energy directly from a spinning black
hole.

During the next several years, one can hope to double the number of
black holes with spins measured via the continuum-fitting method.  It
will be equally important to improve the quality of each measurement,
largely by obtaining more accurate measurements of the parameters $D$,
$i$ and $M$, but also by making methodological advances and by
pursuing more advanced GRMHD and atmosphere models of thin disks.  The
payoff for this effort will be the widening applications of these spin
data to problems in astrophysics and physics.

Especially important will be the possibility of validating both the
continuum-fitting and Fe-line methods by comparing spin results
obtained for individual stellar black holes.  The validation of the
Fe-line method is particularly important because it is the most direct
approach to measuring the spins of AGN.  Presently, concordant results
are being obtained using these two leading methods
\citep[e.g.,][]{ste+2011,ste+2012b,gou+2011, fab+2012,rey+2013}.

Two other methods for measuring the spins of stellar black holes
appear promising, namely, via X-ray polarimetry
\citep{dov+2008,lil+2009,sch+2009} and high-frequency quasiperiodic
oscillations \citep{tor+2005,rem+2006,bel+2012}.  The former method is
stymied by a lack of data, and the latter by the lack of an adequate
model.  It is reasonable to hope that the mass and spin data in Table
1 will assist in identifying the appropriate physical model for the
high-frequency QPOs.  Because spin is such a critical parameter, it is
important to attempt to measure it by as many methods as possible, as
this will arguably provide our best check on the results.  Stellar
black holes are central to this effort because all of the methods of
measuring spin mentioned above can be applied to them.

\begin{acknowledgements}
The authors thank S.~W.~Davis for important input on
Section~\ref{sec:error_atmosph}.  We also thank C.~Brocksopp,
E.~Kuulkers, M.~L.~McCollough, C.~S\'anchez-Fern\'andez and C.~Zurita
for help in preparing Figure~2; J.~Garc\'ia and T.~Fragos for their
comments on a version of the manuscript; R.~Fender for discussions on
MeerKAT; and an anonymous referee for several important criticisms.
JEM was supported in part by NASA grant NNX11AD08G and RN by NASA
grant NNX11AE16G.  JFS was supported by NASA Hubble Fellowship grant
HST-HF-51315.01.
\end{acknowledgements}


\newcounter{BIBcounter} 
\refstepcounter{BIBcounter}

\bibliography{ms2}
\bibliographystyle{apj_8}

\mbox{~}


\end{document}